\documentclass[useAMS,usenatbib]{mn2e}
\usepackage{amssymb}
\usepackage[percent]{overpic}
\usepackage{times}
\usepackage{graphicx}
\usepackage{amsmath}
\usepackage{mathabx}
\usepackage{subfigure}
\usepackage{natbib}
\usepackage{txfonts}
\usepackage{journals}

\newcommand{\HI}{$\rm H~{\scriptstyle I}$}
\newcommand{\Ha}{\ensuremath{\mathrm{H\alpha}}}

\newcommand{\kms}{$\rm km\,s^{-1}$}
\newcommand{\bsnr}{$\rm S/N(4750\,\AA)$}

\title[The S147 dust cloud]
{Mapping the three-dimensional dust extinction toward the supernova remnant S147 -- the S147 dust cloud}

\author[B.Q. Chen et al.]
{B.-Q. Chen,$^{1,2}$\thanks{E-mail:
bchen@pku.edu.cn (BQC); x.liu@pku.edu.cn (XWL).}\thanks{LAMOST Fellow.}
 X.-W. Liu,$^{1,2,3}$\footnotemark[1]
 J.-J. Ren,$^{2,4}$
 H.-B. Yuan,$^5$
 Y. Huang,$^{1,2}$\footnotemark[2]
 B. Yu,$^5$
     \newauthor
 M.-S. Xiang,$^4$\footnotemark[2] 
C. Wang,$^2$
Z.-J. Tian$^2$\footnotemark[2]
and H.-W. Zhang$^2$
\\
$^{1}$South-Western Institute for Astronomy Research, Yunnan University, Kunming, Yunnan 650091, P.\,R.\,China\\
$^{2}$Department of Astronomy, Peking University, Beijing 100871, P.\,R.\,China\\
$^{3}$Kavli Institute for Astronomy and Astrophysics,
Peking University, Beijing 100871, P.\,R.\,China\\
$^{4}$National Astronomy Observatories, Chinese Academy of Sciences, Beijing 100012, P.\,R.\,China\\
$^{5}$Department of Astronomy, Beijing Normal University, Beijing 100875, P.\,R.\,China
 }

\begin{document}

\date{Accepted ???. Received ???; in original form ???}

\pagerange{\pageref{firstpage}--\pageref{lastpage}} \pubyear{2016}
\maketitle
\label{firstpage}

\begin{abstract}
We present a three dimensional (3D) extinction analysis in the region toward the supernova remnant 
(SNR) S147 (G180.0--1.7) using multi-band photometric data from the Xuyi Schmidt Telescope 
Photometric Survey of the Galactic Anticentre (XSTPS-GAC),  2MASS and WISE.  
We isolate a previously unrecognised dust structure likely to be associated with SNR S147. 
The structure, which we term as ``S147 dust cloud'',
is estimated to have a distance $d=1.22 \pm 0.21$\,kpc, consistent with
the conjecture that  S147 is associated with pulsar PSR J0538 + 2817. 
The cloud includes several dense clumps of relatively high extinction that 
locate on the radio shell of S147 and
coincide spatially with the CO and gamma-ray emission features. 
We conclude that the usage of CO measurements to trace the SNR associated MCs is 
unavoidably limited by the detection threshold, dust depletion, 
and the difficulty of distance estimates in the outer Galaxy.  
3D dust extinction mapping may provide a better way to identify and study SNR-MC interactions.
\end{abstract}


\section{Introduction}
\label{introduction}

Supernova (SN) explosion is one of the most energetic events 
in a galaxy. It usually occurs near its maternal molecular cloud (MC). As such,
MCs generally play a critical role in the evolution 
of the supernova remnants (SNRs). There are now $\sim$300 SNRs identified in
our Milky Way \citep{Green2014}. About 70 of them are
known to be or possibly be interacting with MCs \citep{Jiang2010,Chen2014b}.
And SNR S147 (G180.0--1.7; hereafter referred to as S147), is claimed to be 
not one of those MC-interacting SNRs \citep{Huang1986, Jeong2012}.

S147 is one of the most evolved SNRs in the Milky Way. It falls near 
the Galactic anticentre, at Right Ascension  
$\alpha$=05h39m00s and Declination $\delta=+7\degr50^{\prime}0^{\prime\prime}$, 
and has a large angular size of about 200$^{\prime}$. It shows a complex network of
long filaments in the optical \citep{vandenBergh1973,Drew2005} and radio 
\citep{Sofue1980, Kundu1980, Xiao2008} bands. The pulsar PSR J0538+2817, located near the 
center of S147, is suggested to be plausibly associated with S147 \citep{Anderson1996,Ng2007}.
The pulsar is estimated to have an age of 3 $\pm$ 0.4 $\times$ 10$^4$\,yr \citep{Kramer2003}. 
X-ray emission has been detected from the pulsar by ROSAT All Sky Survey \citep{Sun1996}.
In gamma-ray, an extended source is found to coincide with S147 \citep{Katsuta2012}.
The $V$-band extinction of S147 is estimated to be $A_V$ = 0.7 $\pm$ 0.2\,mag \citep{Fesen1985}
from optical spectroscopy. \citet{Dincel2015} find that HD\,37424, possibly the 
pre-supernova binary companion of the pulsar, has an extinction
of $A_V$ = 1.28 $\pm$ 0.06\,mag.
The distance of S147 given in the literature varies between 0.6 and 1.6\,kpc (Table~1).
From the parallax of the pulsar, S147 is estimated to be at $1.30^{+0.22}_{-0.16}$\,kpc 
\citep{Chatterjee2009}, more remote than estimated from the 
$\Sigma$-$D$ relation and by optical absorption analysis.

\begin{table}
\caption{An updated list of distance estimates 
for S147 based on Table 1 of \citet{Dincel2015}. 
(R$^{l}$) denotes the lower limit radius of the SNR.}
\begin{center}
\begin{tabular}{lll}
\hline
\hline
Distance (kpc) & Method & Reference \\\hline
1.6$\pm$0.3		 & $\Sigma-$\text{D}          &  {\citet{Sofue1980}} \\
0.8--1.37		 & R$^{l}$, $\Sigma-$\text{D} &  {\citet{Kundu1980}} \\
0.6			 & R$^{l}$          &  {\citet{Kirshner1979}} \\
0.9			 & $\Sigma-$\text{D}          &  {\citet{Clark1976}} \\
0.8$\pm$0.1		 & $A_V$              &  {\citet{Fesen1985}} \\
1.06			 & $\Sigma-$\text{D}         &  {\citet{Guseinov2004}} \\
$<0.88$			 & High Vel Gas     &  {\citet{Sallmen2004}} \\
1.2			 & Pulsar DM        &  {\citet{Kramer2003}} \\
1.47$^{+0.42}_{-0.27}$	 & Pulsar Plx       &  {\citet{Ng2007}} \\
1.3$^{+0.22}_{-0.16}$	 & Pulsar Plx       &  {\citet{Chatterjee2009}} \\
1.333$^{+0.113}_{-0.112}$ & OB runaway star & {\citet{Dincel2015}} \\
1.22$\pm$0.21 & 3D dust mapping & this work \\
\hline
\end{tabular}
\end{center}
\end{table}

\citet{Phillips1981} have detected low-velocity interstellar 
CO absorption features in the UV spectrum toward HD\,36665 ($d$=0.9\,kpc, once
believed to lie behind S147). They claim that the 
SNR is evidently expanding into an inhomogeneous interstellar medium. 
\citet{Sallmen2004} have studied the interstellar Na\,{\sc i}  and Ca\,{\sc ii} 
absorption lines in the high-resolution spectra of three stars at different distances in sight-lines 
towards S147. They find that two of the more distant stars, HD\,36665 ($d$ = 880\,pc) 
and HD\,37318 ($d$ = 1380\,pc), possess complex absorption features over intermediate velocities,
while the nearest, foreground star, HD\,37367 ($d$=360\,pc), exhibits none of those features.
Does S147 physically associated with any MCs? In general, a SNR-MC association is normally 
established with two types of study: 1) observation and  detection of 
OH 1720\,MHz masers \citep{Frail1996, Green1997}; and 
2) sub-millimeter/millimeter molecular line
and infrared emission line observation and analysis \citep{Huang1986, Jeong2012}. 
Given the weakness of the maser emission, many more SNR-MC associations may yet
remain to be revealed \citep{Hewitt2009}.
In such cases, CO is a practical tracer of MCs  
commonly used to trace the SNR-MC interaction. 
\citet{Huang1986} present an early CO-line survey toward 26 outer 
Galactic SNRs with Galactic longitude $l$ between 70\degr\ and 210\degr.
For S147, they do detect a weak cloud at a peak velocity around 4\,km\,s$^{-1}$.
They suggest that the cloud is probably in front of the SNR. Similarly, although
a most recent CO survey for  Galactic SNRs with $l$ between 60\degr\ and
190\degr\ by \citet{Jeong2012} also find some MCs 
at velocities between $\sim$ $-$14 and $+$5 \kms, they believe none 
of those clouds are with the S147. 

However, \citet{Katsuta2012} have reported a spatially extended gamma-ray
source  likely to be associated with SNR S147, 
suggesting possible interactions between the MCs and the SNR blast wave.
The results of \citet{Huang1986} and \citet{Jeong2012} 
with regard to S147 are thus open to further discussion.  
In the direction of Galactic anti-center, 
the kinematic distance can hardly be  
determined from the local standard of rest (LSR) velocity, 
and thus be used to establish the possible association of a SNR with a MC.
It is thus unclear whether the MCs
found by \citet{Huang1986} and \citet{Jeong2012} are associated with S147 or not. 
In general, the strong CO $J=$1--0 emission is a good tracer of MCs. However the low density 
regions of a MC could be below the column density threshold 
required for a detection \citep{Goodman2009, Chen2014}. 
There is also evidence that there is a substantial fraction
of ``dark'' molecular gas not traced by CO emission \citep{Planck2011,Chen2015}. In such cases, 
tracing the molecular gas using dust observation, in particular 
via the optical and/or the near-infrared dust extinction, 
seems to be a good, essential alternative way \citep{Goodman2009}.  
However, considering the traditional 
extinction maps lack the information of distance, 
obtaining distance-dependent three-dimensional (3D) extinction maps remains a challenge 
for this purpose. 

With the availability of large amounts of multi-band photometric 
as well as spectroscopic data, thanks to a number of either already completed or 
still on-going large area or all-sky surveys, deriving 3D dust extinction maps for 
large sky areas has become an acknowledged  aspect of Galactic studies that has
received increasing attention in the recent years
\citep{Chen2013,Chen2014, Schultheis2014, Green2015, Hanson2016}.    
In this work, we improve the 3D dust mapping by \citet{Chen2014} 
for the Galactic anticentre area and use the result to differentiate 
the foreground and background dust extinction of S147. 
The work reveals a large dust cloud, parts of which are possibly 
associated with S147, which we term as the ``S147 dust cloud.''
This paper is one of a series of  three papers
on the study of S147 based on the data of the Xuyi Schmidt Telescope
photometric survey of the Galactic Anticentre (XSTPS-GAC, 
\citealt{Liu2014,Zhang2013,Zhang2014}) as well as of the  
LAMOST spectroscopic Survey of the Galactic Anticentre 
(LSS-GAC, \citealt{Liu2014,Liu2015}). In the other two papers, we 
will present the S147 kinematics (Ren et al., in preparation), as well as the S147 dust properties
including possible presence/absence of diffuse interstellar bands (DIBs) 
in the S147 dust cloud (Chen et al., in preparation).   

In this paper, we first introduce our data set in Sect.~2.
We describe in Sect.~3 our method to determine the extinction 
and distance of individual stars, as well as to derive the 3D 
dust extinction distribution. In Sect.~4 we present the results. We compare
the morphology of the newly discovered S147 dust cloud with 
images of gas emissions from the radio to the gamma-ray 
in Sect.~5. We finish with Sect.~6 with the summary.  

\section{Data}

Our work uses the optical  photometry of the
multi-band CCD photometric survey of the Galactic 
Anticentre with the Xuyi 1.04/1.20m Schmidt Telescope  
(XSTPS-GAC; \citealt{Zhang2013,Zhang2014,Liu2014}).
XSTPS-GAC collected data during the fall of 2009 and the spring of 2011,
using the Xuyi 1.04/1.20\,m Schmidt Telescope equipped with a 4k$\times$4k CCD camera, 
operated by the Near Earth Objects Research Group of the Purple Mountain Observatory.
The survey took images in SDSS $g$, $r$ and $i$ bands. 
In total, XSTPS-GAC archives approximately 100 million stars down to a
limiting magnitude of about 19 in $r$ band ($\sim$ 10$\sigma$). 
The astrometric accuracy is about 0.1$^{\prime\prime}$
and the global photometric accuracy is about 2\,per\,cent \citep{Liu2014}.
The total survey area of XSTPS-GAC is close to 7,000\,deg$^2$,
including an area of $\sim$ 5, 400\,deg$^2$ centered on the Galactic anticentre,
from $\alpha$ $\sim$ 3 to 9\,h and $\delta$ $\sim~-$10\degr\ to $+60\degr$, 
an extension of about 900\,deg$^2$ to the M31/M33 area and the bridging fields 
which connect the two areas.

The infrared photometry from the Two Micron All Sky Survey (2MASS; \citealt{Skrutskie2006}) 
and the Wide-field Infrared Survey Explorer (WISE; \citealt{Wright2010}) are also adopted in 
the work, in order to break the degeneracy between 
effective temperature (or intrinsic colours) and extinction \citep{Berry2012, Chen2014}.
2MASS is an all-sky survey in $J$, $H$ and $K_{\rm s}$ bands,  
centered at 1.25, 1.65 and 2.16 $\mu$m, respectively. 
The survey was conducted with two 
1.3-m telescopes located respectively at Mount Hopkins, 
Arizona and Cerro Tololo, Chile, in order to provide full coverage of 
both the northern and southern skies.
For point-sources, the limiting magnitudes of 2MASS (S/N = 10) are approximately
15.8, 15.1 and 14.3\,mag in $J$, $H$ and $K_{\rm s}$ respectively.
The systematic photometric uncertainties of 2MASS are estimated to be 
smaller than 0.03\,mag, while the astrometric 
uncertainties are less than\,0.2$^{\prime\prime}$.

WISE is an all-sky survey in four infrared bands, $W1$, $W2$, $W3$ and $W4$, centered 
at 3.4, 4.6, 12 and 22 $\mu$m, respectively. 
The survey was conducted from a 40\,cm telescope on board the satellite.
Instead of the WISE All-Sky Release Catalog adopted by \citet{Chen2014}, in the current work we use
the AllWISE Source Catalog which contains positions, proper motions, 
photometry and ancillary information of 748 million objects detected in the deep, 
coadded AllWISE Atlas Images \citep{Kirkpatrick2014}. 
The adopted AllWISE Source Catalog represents a significant improvement 
over the WISE All-Sky Release Catalog. It is significantly more sensitive in $W1$ and $W2$ bands, 
almost doubled, since all usable images 
from the whole WISE survey have been combined together.
Furthermore, the AllWISE photometry is less biased and source astrometry is also improved.
All sources in the AllWISE Source Catalogue have a 
measured S/N greater than 5 in at least one of the four bands. 
The AllWISE Source Catalog is 0.3--0.5 magnitudes fainter 
than the WISE All-Sky Release Catalog, with
$95$\,per\,cent source completeness at 17.1, 15.7, 11.7 and 7.7\,mag in
$W1$ to $W4$ band, respectively.

\section{Method}

\begin{figure}
  \centering
  \includegraphics[width=0.48\textwidth]{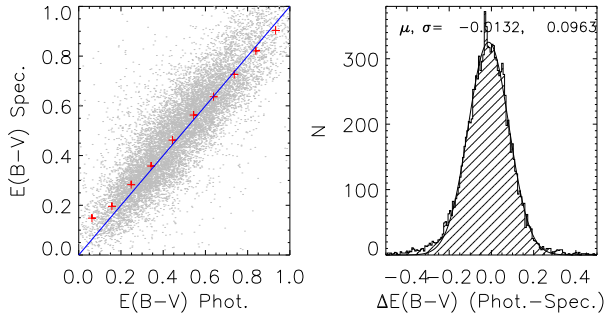}
  \caption{Comparisons of extinction values derived in  
the current work and recommended values presented in the LSS-GAC DR2 \citep{Xiang2016}.
The red pluses in the left panel represent median values in the individual bins.
A blue straight line denoting complete equality is also overplotted to guide the eyes. 
In the right panel, the black curve is a Gaussian fit 
to the distribution of differences of values.}
  \label{psebv}
\end{figure}

We select stars in a region centered at S147,
between $177\degr\ <l<183\degr$ and $-5\degr\ < b < 1\degr$.
The determination of extinction for individual stars is similar as that of \citet{Chen2014}.
\citet{Chen2014} cross-matched the photometric catalogue of XSTPS-GAC with those of 
2MASS and WISE, and applied spectral energy distribution (SED) fitting to the data.
They obtained the best-fit values of extinction and distance for over
13 million stars and constructed a 3D extinction map of the Galactic anticentre. 
In the current work, we implement several improvements 
to  the method of \citet{Chen2014}, described as follows. 
\begin{enumerate}
\item In creating the library of standard stellar loci, \citet{Chen2014} adopted the observed apparent 
colours of stars from their ``high-quality reference sample of (essentially) zero extinction'' directly.
The extinction of those stars are indeed 
rather small ($A_r < 0.075$\,mag as given by the 2D map of \citealt{Schlegel1998}).
Yet the reference stellar loci thus derived should still be slightly redder than they actually are. Consequently, 
the resultant extinction values may be slightly underestimated. 
\citet{Yuan2015} find that the $E(B-V)$ reddening values of \citet{Chen2014} are 
systematically smaller than their values derived from spectroscopic data by about 0.06\,mag. 
This systematical difference is partly caused by the usage of a stellar locus library
 that has not been corrected for reddening by Chen et al. (2014a). 
In the current work, we have dereddened stars 
in the ``high-quality reference sample of (essentially) zero extinction'' and recreated the
library of standard stellar loci. The colour differences between the new stellar locus and the old ones 
are however found to be no larger than 0.01\,mag on average.
\item \citet{Chen2014} calculated extinction and distances of 
individual stars that are detected in  all eight bands 
($g,~r,~i,~J,~H,~K_{\rm s},~W1$ and $W2$). A large number 
($\sim$ 10\,per\,cent) of stars were consequently excluded as 
many of them were detected only in some of the infrared bands. 
In the current work, we have accepted all stars detected in  
all the three XSTPS-GAC optical bands ($g,~r,$ and $i$) 
and at least in any  two infrared bands (2MASS $J,~H,$
and $K_{\rm s}$  or WISE $W1$ and $W2$). The new 
criteria enlarge  the sample by 13\,per\,cent.
\end{enumerate}

Finally, we select stars with minimum $\chi^2 <2$ and exclude those
flagged as giant stars   following the procedure of \citet{Chen2014}. 
This leads us to 161,642 stars with extinction values determined.
The results are compared with those published in the second 
data release of value-added catalogue of LSS-GAC (LSS-GAC DR2; \citealt{Xiang2016}).
In LSS-GAC DR2, the ``recommended'' values of extinction were derived 
using the so-called ``star pair'' method \citep{Yuan2015}.
We select from LSS-GAC DR2 stars with spectral signal 
to noise ratios (S/Ns) in blue band (4750\AA) larger than 15 
(`{\rm snr\_b}' $> 15$) and unaffected by bright lunar light ( `moondis' $> 60$). 
In total there are 11,861 stars in common with the current sample. 
The comparison between our extinction values and those from \citet{Xiang2016} 
is shown in Fig.~\ref{psebv}. Our extinction values are in excellent agreement 
with the results from spectroscopic data. After corrected for the extinction of 
stars used to create the reference stellar
loci, the systematic differences between our photometric extinction 
values and those based on spectroscopic analysis, are reduced to   
only about 0.01\,mag, which is much lower
than that found by \citeauthor{Yuan2015} (2015; 0.06\,mag) and thus can be entirely ignored. 
The differences have a dispersion just shy of 0.1\,mag. 

\begin{figure}
  \centering
  \includegraphics[width=0.48\textwidth]{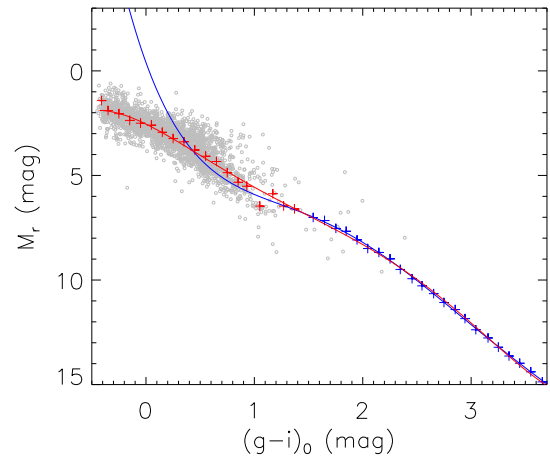}
  \caption{Photometric distance calibrated with the spectroscopic data. The grey dots
show absolute magnitude $M_r$ versus intrinsic colour $(g-i)_0$ for the individual stars. 
The red pluses are median values of the individual bins. The blue curve represents 
the \citet{Ivezic2008} relation for metallicity [Fe/H]=$-$0.15\,dex, and blue pluses mark 
selected values of the curve. The red curve is a  5-order polynomial fit of 
to red and blue pluses.}
  \label{disc}
\end{figure}

\begin{figure}
  \centering
  \includegraphics[width=0.48\textwidth]{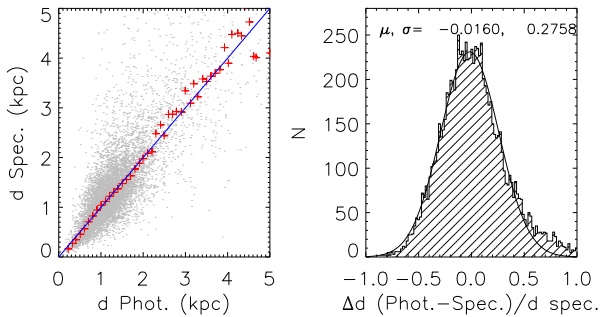}
  \caption{Comparison of distances derived in the
current work and those from \citet{Xiang2016}.  
The red pluses in the left panel represent median values in the individual bins.
A blue straight line denoting complete equality is overplotted 
to guide the eyes. In the right panel, the black curve is a 
Gaussian fit to the distribution of distance differences.}
  \label{psdis}
\end{figure}

\citet{Chen2014} calculated distances of the individual stars using the \citet{Ivezic2008}
photometric parallax relation, assuming that all stars have a metallicity of [Fe/H] = $-$0.2\,dex.
Since the \citet{Ivezic2008} relation are based on globular clusters, which are usually
of old ages, the relation may not be suitable for blue stars of young age of interest here. This effect is clearly  
visible in Fig.~3 of \citet{Chen2014}. In the case of young cluster, M35, 
its distance yielded by blue stars [$(g-i)_0 < 0.5$\,mag] were clearly overestimated.
Those blue young stars are still on the main sequence, 
while the  \citet{Ivezic2008} relation is applicable  
only to stars of old ages (turnoff and sub-giant stars).
In the current work, we solve this problem by recalibrating the 
photometric distance relation using the spectroscopic distances
from LSS-GAC DR2 \citep{Xiang2016}.
Those spectroscopic distances are based on  
absolute magnitudes directly estimated from LAMOST spectra 
with the KPCA regression method using the LAMOST-Hipparcos
training set \citep{Xiang2016b}. An obvious advantage of this approach is that absolute 
magnitudes thus estimated directly from LAMOST spectra are independent of any stellar model 
atmospheres and atmospheric parameters. The results are thus expected to suffer from minimal systematics. 
A comparison of the spectroscopic distances with the Gaia-TGAS parallaxes \citep{Gaia2016}
 shows that the typical uncertainties of thus deduced absolute magnitudes are only 0.3\,mag 
for high quality spectra [\bsnr $>$ 50], corresponding to a distance error of 15\,per\,cent \citep{Xiang2016b}. 

To (re-)calibrate the photometric parallax relation using spectroscopic distances, 
we select stars from LSS-GAC DR2 using the criteria: `snr\_b' $> 50$ for
high blue band spectral S/Ns, log\,$g > 3.8$\,dex in order to exclude giant stars,
and  `moondis' $> 30$ for avoiding bright lunar light pollution.
This yields 2,356 stars in common with our current sample. Absolute magnitudes in $r$-band $M_r$ of
those stars from LSS-GAC DR2 are plotted against intrinsic colours $(g-i)_0$ from
the current work. Also overplotted in the Figure is the \citet{Ivezic2008} relation assuming 
metallicity [Fe/H] = $-$0.15\,dex\footnote{We have obtained a peak metallicity [Fe/H] = $-$0.15\,dex for
all LSS-GAC stars in this region.}. Since we have adopted a relatively high S/N cut, most  of the stars 
are blue ones with a bright apparent magnitude. 
The \citet{Ivezic2008} relation seems to yield satisfactory results for 
stars redder than  $(g-i)_0 > 0.5$\,mag. For blue stars, the 
relation gives brighter absolute magnitudes than those
from LSS-GAC  DR2, leading to overestimated distances, 
as seen in the case of M35 in \citet{Chen2014}.  
To construct a new photometric parallax relation that is also valid for blue stars,
a 5-order polynomial is used to fit $M_r$ as a function of $(g - i)_0$ colour,
using the binned median values of spectroscopic absolute 
magnitudes from LSS-GAC DR2. Since only very few red stars
 are available from LSS-GAC DR2, stars of  $(g-i)_0 > 1.5$\,mag, 
 we use simulated data for those redder colours created from the \citet{Ivezic2008} relation,
assuming [Fe/H]=$-$0.15\,dex. The fit yields,
\begin{eqnarray}
M_{r} & = &  2.547 +2.533(g-i)_0 +1.546(g-i)_0^2 \nonumber \\
          &   &   -1.578(g-i)_0^3 +0.581(g-i)_0^4 -0.0679 (g-i)_0^5.
\end{eqnarray}
The relation is valid for stars over a wide colour range of  $-$0.5 $<  (g - i)_0 <$ 3.7\,mag. 
This relation is then applied to
all star in our sample, with distances of the individual stars calculated 
using the standard relation,
\begin{equation}
d=10^{0.2(r-A_r-M_r)+1}.
\end{equation}
In Fig.~\ref{psdis}  we compare the resultant distances with those 
from LSS-GAC DR2 for common stars of a LAMOST spectral \bsnr $>$ 15. 
Photometric distances from the newly constructed relation  
are in good agreement with the spectroscopic values. 
The differences are well fitted by a
Gaussian. The systematic difference
between our new photometric distances and those LSS-GAC spectroscopic values  is
only 1\,per\,cent, with a dispersion of 28\,per\,cent. 
Considering that uncertainties of spectroscopic  distances for stars of a spectral \bsnr $> $ 20
are about 20\,per\,cent \citep{Xiang2016b},
the uncertainties of the photometric distances
derived  in this work should also be about 20\,per\,cent.

\begin{figure}
  \centering
  \includegraphics[width=0.48\textwidth]{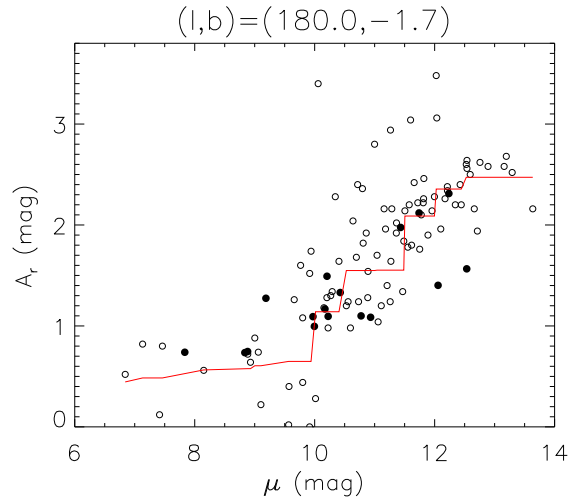}
  \caption{Mapping the 3D extinction for an example field at 
$(l, b) = (180.0\degr, -1.7\degr)$ of size 0.1\degr\ $\times$ 0.1\degr.  
The open and filled circles represent $r$-band extinction and distance  
values derived in the current work and those from LSS-GAC DR2, 
respectively.  The red line 
shows the inferred maximum likelihood extinction as a function of distance.}
  \label{slos}
\end{figure}

With reddening values and distances deduced for individual stars in our sample, 
the 3D dust distribution toward S147 is then mapped.
The procedure is similar to that of \citet{Green2014}.
We first divide our sample into subfields of size $6^\prime$ $\times$ $6^\prime$.
This resolution is chosen to balance between high angular resolution
and sufficient number of stars in each pixel (subfield)  in order
to obtain a robust  extinction versus distance 
relation for the pixel. We have also tried alternative resolutions. For higher resolutions, a large fraction
of pixels have not enough stars to obtain a robust extinction profile for the pixel. For lower
resolutions, while we obtain similar results,  some detailed features 
of the resultant dust distribution are washed out.
For each pixel, we derive an $r$-band extinction profile $A_r(\mu)$, using the
extinction values and distances thus obtained for individual stars, 
where $\mu$ is the distance module given by $\mu = 5{\rm log}\,d - 5$.
We parameterise $A_r(\mu)$ by a piecewise linear function, given by,
\begin{equation}
A_r(\mu) = \Sigma_{0}^{\mu} (\Delta A^i_r),
\end{equation} 
where $\Delta A_r^i$ is the local extinction in each distance bin and $i$ the index of distance bin.
The size of each distance bin is set to $\Delta \mu = 0.5$\,mag.
A MCMC analysis is performed to find the best set of 
$\Delta A^i_r$ that maximises the likelihood defined as,
\begin{equation}
L = \Pi_{n=1}^{N} \frac{1}{2\pi \sigma_n}{\exp}(\frac{-(A^n_r-A^n_r(\mu))^2}{2\sigma^2_n}),
\end{equation}
where $n$ is index of stars in the pixel, 
$A^n_r$ and $A^n_r(\mu)$ are respectively extinction  derived in the current
work and that simulated from Eq.~(3) of the star, and $\sigma_n$ is the 
total uncertainty of the derived extinction and distance, given by
$\sigma_n=\sqrt{\sigma^2(A_r)+(A_r\frac{\sigma(d)}{d})^2}$ 
\citep{Lallement2014}. Note, however, that the error $A_r\frac{\sigma(d)}{d}$ resulting 
from the distance error is only an approximation for the assumption that the 
opacity is constant along the line of sight.

An example for the line of sight ($l,b$)=(180\degr, $-$1.7\degr), near the centre of S147, 
is shown in Fig.~\ref{slos}.  
There is little dust in distance bins of $\mu < 10\,$mag.
A sudden `jump' in extinction, by about $1$~mag $A_r$, is clear visible
at $\mu \sim 10$\,mag, indicating the presence of a dust cloud at this distance.
The extinction values and distances from LSS-GAC 
DR2 are also overplotted in the Figure as filled circles. They agree well with
the extinction profile derived here. 

\section{The S147 dust cloud}   

\begin{figure*}
  \centering
  \includegraphics[width=0.78\textwidth]{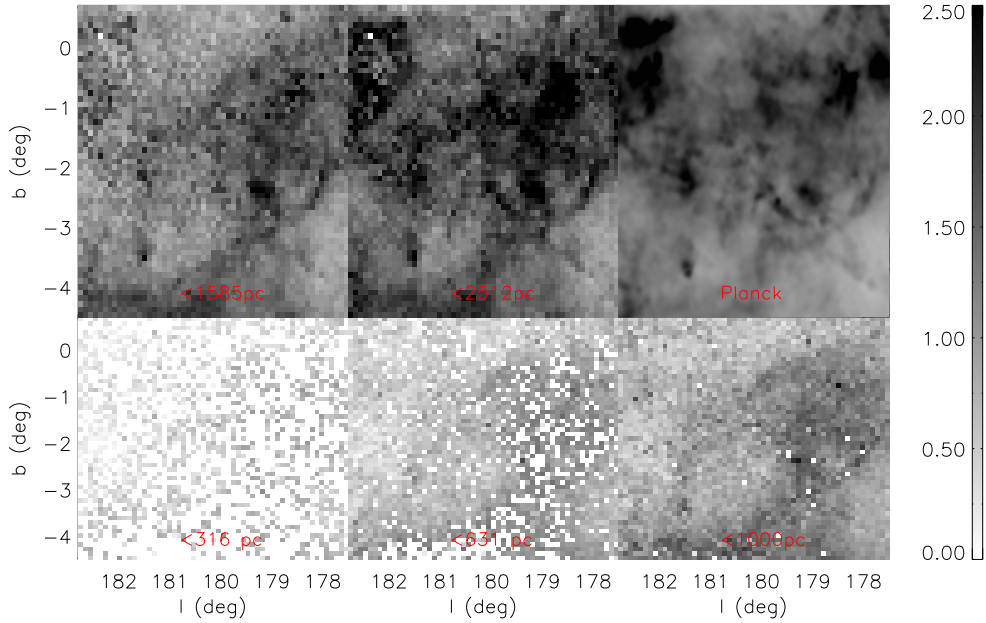}
  \caption{2D extinction maps integrated to various distances in the direction of  S147. 
X-axis denotes Galactic longitude while y-axis Galactic latitude. 
The distances, labelled in in each panel, increase from bottom left to top right.
The top right panel shows the 2D extinction map from \citet{Planck2014}
 for comparison.}
  \label{intdust}
\end{figure*}

\begin{figure*}
  \centering
  \includegraphics[width=0.78\textwidth]{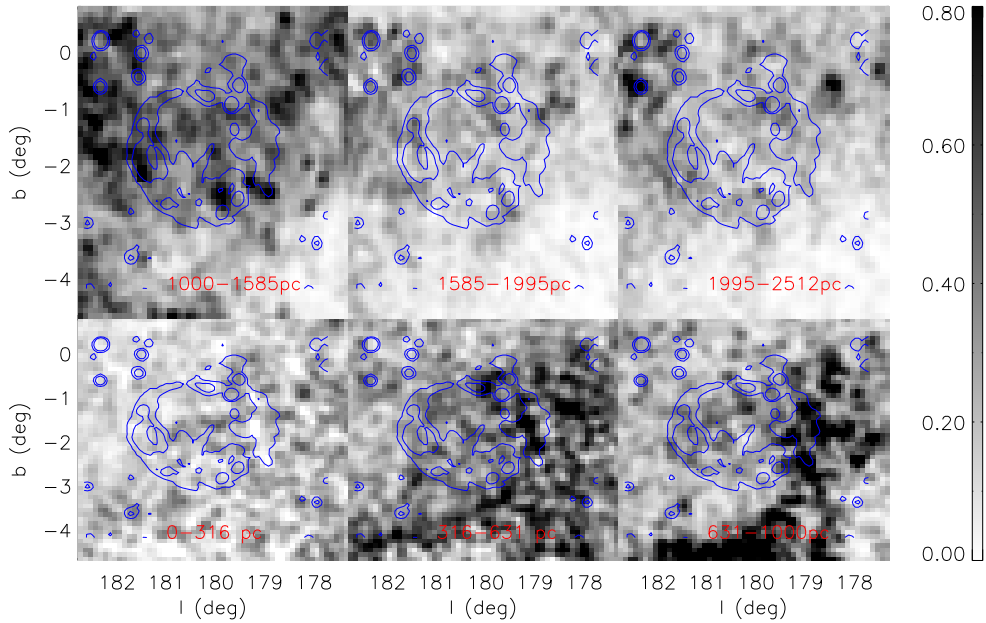}
  \caption{Distributions of dust extinction in individual distance bins in the direction of  S147.
The distance ranges of the bins are labelled in the individual panels. 
The x-axis denotes the Galactic longitude and the y-axis the Galactic latitude.  
The blue contours overplotted in all panels 
are total intensity map of 6\,cm radio emission of ionised gas  of S147 from \citet{Xiao2008}.}
  \label{dust}
\end{figure*}

\begin{figure*}
  \centering
  \includegraphics[width=0.68\textwidth]{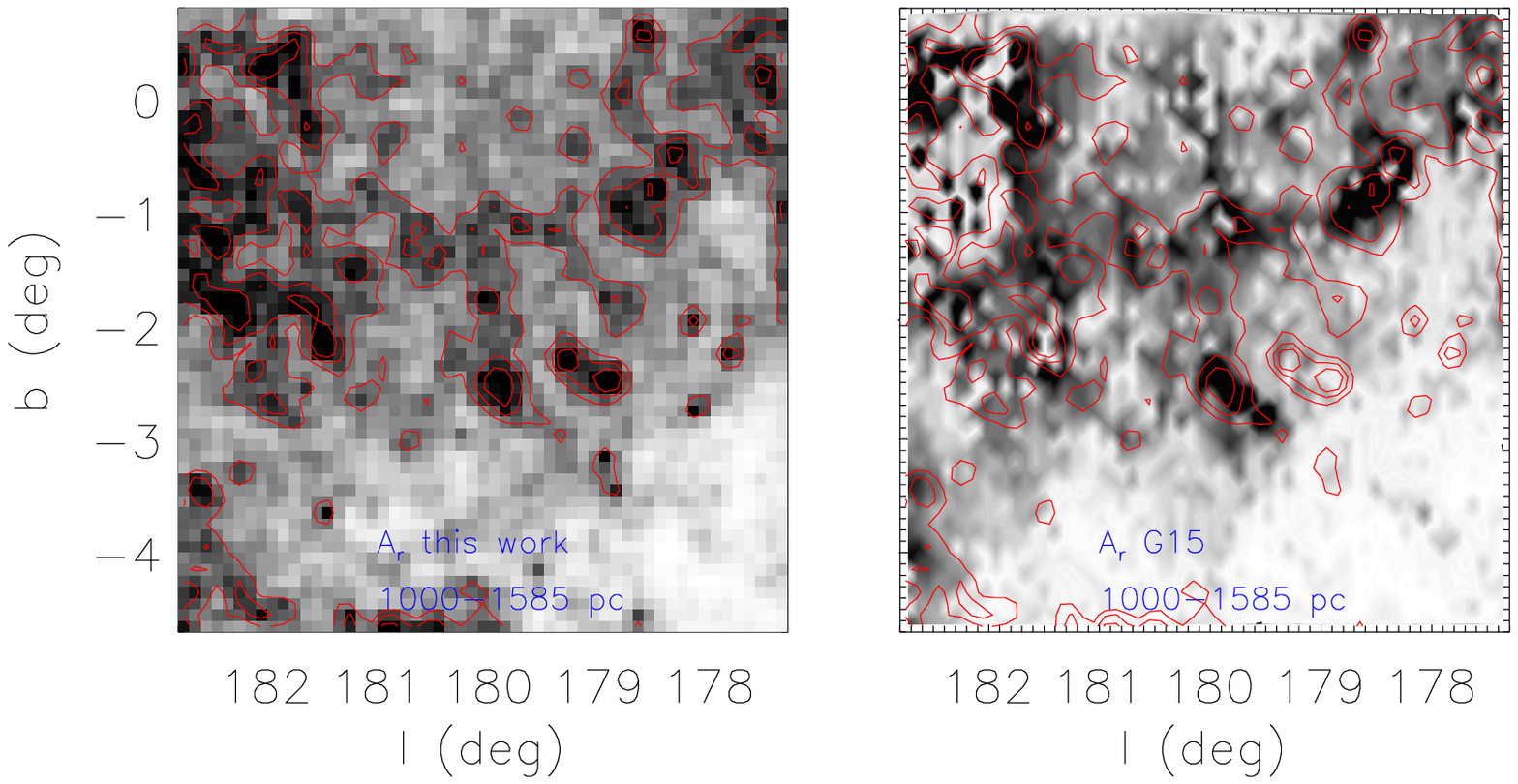}
  \caption{Distribution of dust extinction in the distance bin 1000 $<~d~<$ 1585\,pc 
in the direction of  S147 obtained in the current work (left panel) and that from
\citet[right panel]{Green2015}.  Red contours in each panel show the dust extinction distribution 
derived in the current work.}
  \label{ps1}
\end{figure*}

On the basis of the resulted reddening profiles of all  pixels, 
we produce a 3D map of dust extinction in an area around S147, 
of $177.5\degr\ < l < 182.5\degr$ and $-4.5\degr\ < b < 0.5\degr$.  
To help visualise variations of  dust extinction as a function of distance,  we 
plot in Fig.~\ref{intdust} 2D $r$-band extinction distributions 
integrated to selected  distances from the Sun. In general, the extinction values
increase with distances for all pixels, but the growth rates vary from pixel to pixel, 
suggesting highly inhomogeneous and clumpy 
dust distributions  along the lines of sight.
For the two maps of nearest distances ($d < $ 631\,pc)  plotted in Fig.~\ref{intdust}, 
there are not enough stars for some of the 
pixels due to the limited volume and the bright saturation limit of XSTPS-GAC  ($\sim 9$\,mag). 
As one approaches a distances of 1.0\,kpc, one encounters a large dust cloud of $l$
between 178\degr\ and 180\degr. This is parts of the Taurus-Perseus-Auriga cloud. 
At further distances ($d \lesssim $ 1.6\,kpc), a  
structure of a significant amount dust is seen spatially coincident with S147. 
An addition dust cloud of $l$ around 182\degr\ and $b$ around 0\degr\ is then 
found at the furthest distances ($d \lesssim 2.5$\,kpc). 
Finally, also plotted in Fig.\,5 is the Planck 2D dust extinction map of the area, 
deduced from dust  far-infrared thermal emission \citep{Planck2014}, 
which can be compared with our map integrated to 2.5\,kpc. 
Both maps show similar dust features, suggesting that the dust 
distribution obtained in the current work is robust.

To highlight dust features in different distance slices, 
we plot in Fig.~\ref{dust} the distributions of dust extinction 
in distance bins $d<316$\,pc, $316<d<631$\,pc, $631<d<1000$\,pc, 
$1000<d<1585$\,pc, $1585<d<1995$\,pc, and $1995<D<2512$\,pc, 
respectively.  For each distance bin, the extinction 
map has been smoothed with a Gaussian kernel of 0.3\degr\ full width at half maximum (FWHM).
The maps reveal the fine structures of dust distribution in those distances bins.
Also overplotted in blue contours of intensity flux of S147 
at 6\,cm radio wavelength \citep{Xiao2008}, 
delineating the distribution of ionised gas of S147.  
The dust seen spatially coincident with the ionised gas of S147 is distributed in distance bin 
$1000< d< 1585$\,pc, while the foreground dust of $l$ between 178\degr\ and 180\degr\
is located at distances $d< 631$\,pc and the background dust of $l$ around 182\degr\ and 
$b$ around 0\degr\ is located in bin $1995 < d < 2512$\,pc. We note that
although some features are seen in distance bin $631 < d < 1000$\,pc, 
that may be attributed to the foreground dust cloud, they are actually artefacts 
caused by insufficient numbers of stars in some pixels 
at the selected spatial resolution of 0.1\degr. 
If one adopts a lower resolution, such as 0.2\degr, 
one sees the foreground dust falls entirely $d < 631$\,pc.  

There is a large dust cloud at distances between $1000<d<1585$~pc
in the direction of S147. Within the 6cm emission area of S147, parts of the dust cloud, 
which has a morphology spatially coincident with the ionised gas of S147,
are referred to as the ``S147 dust cloud''. 
There are three dense clumps of relatively high extinction 
($\Delta A_r $ $\sim$ 1\,mag) in S147 dust cloud, which are located
on the radio shell of S147,
respectively at  $l ~\sim$ 181.3\degr\ and $b~\sim~-$2.2\degr, 
$l ~\sim$ 179.9\degr\ and $b~\sim~-$2.7\degr\ and 
$l ~\sim$ 179.0\degr\ and $b~\sim~-$2.5\degr.  
In Fig.~\ref{ps1} we compare our derived dust distribution at distances between $1000<d<1585$~pc
to that of \citet{Green2015}, who present a 3D map of the interstellar dust reddening
deduced using the Pan-STARRS 1 and 2MASS photometry. Values of
$E(B-V)$ of \citet{Green2015} are converted to $A_r$ using the coefficient given by \citet{Schlafly2011}.
 In general, the agreement is good.
The morphology of the S147 dust cloud derived here shows quite similar features 
as that of \citet{Green2015}, except for the dense clump at $l ~\sim$ 179.0\degr\ and $b~\sim~-$2.5\degr.
This clump could be parts of the foreground dust cloud that has been wrongly placed 
at an artificially larger distance due to the uncertainties in distance estimation and
the insufficient numbers of stars in those pixels (See also in Fig.~\ref{slice}).
A detailed examination of the S147 dust cloud, 
including a precise estimate of its distance as well as of  mass, is presented below.

\subsection{Distance}

\begin{figure*}
  \centering
  \includegraphics[width=0.88\textwidth]{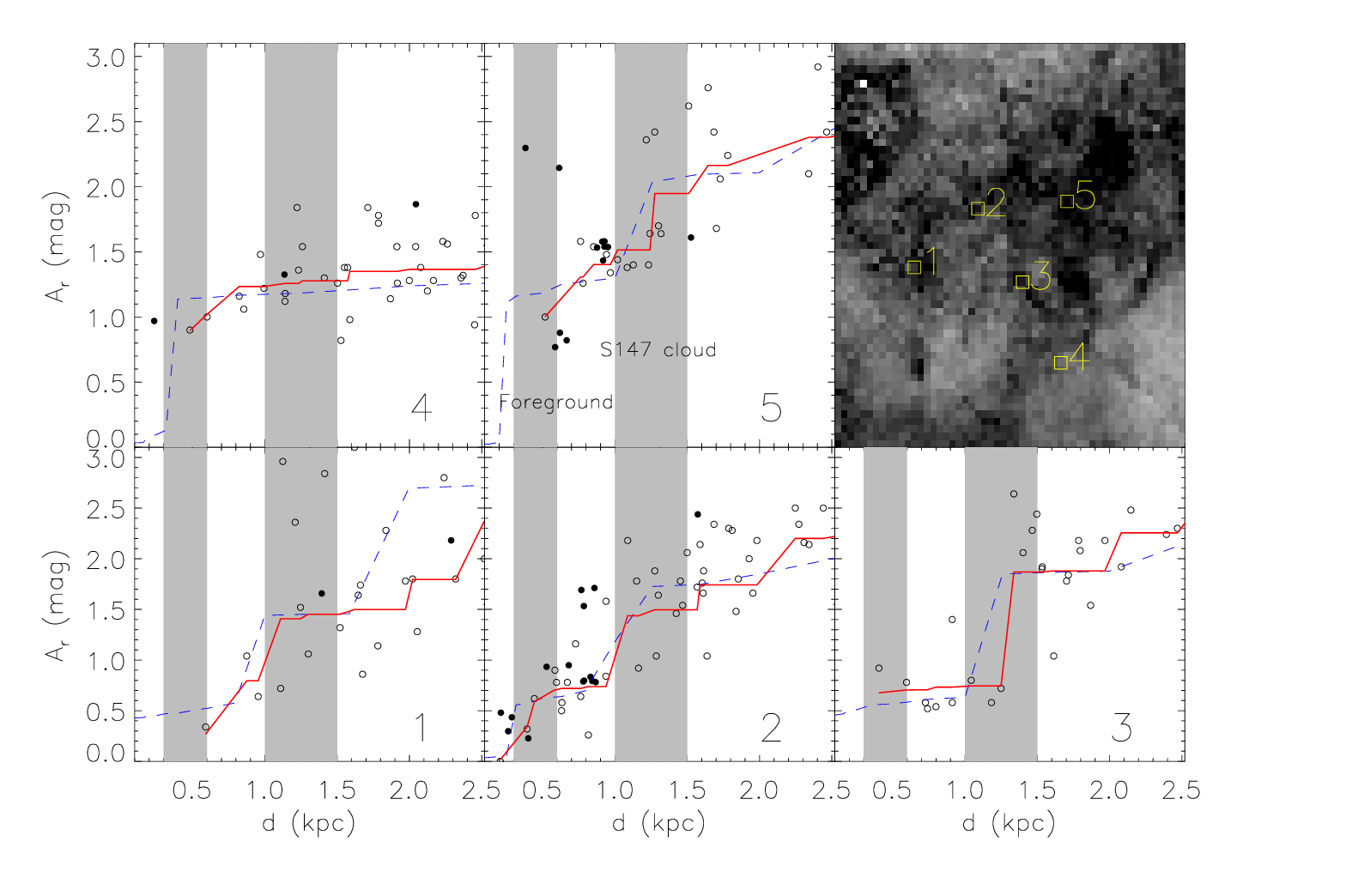}
 \caption{Extinction values as a function of distance of stars in five select lines of sight.  
Our extinction map integrated to 2.5\,kpc is shown in the top-right panel, with
the selected lines of sight marked.  
The grey scale corresponds to a 0--2.5\,mag  range of extinction in $r$-band, $A_r$.
The other five panels, marked by the number of sight line~1 to 5, show values of 
extinction as a function of  distance for stars within the $0.1\degr \times 0.1\degr$ 
pixel area of the sightline. The open and
filled circles correspond to stars in our current 
sample and those available from LSS-GAC DR2, respectively.
The vertical shadows delineate  the distance ranges of the foreground dust cloud at about 0.5\,kpc
and the S147 cloud at about 1.2\,kpc. The red solid lines are best-fit extinction-distance relations 
to the data, while  the blue dashed lines are corresponding relations from \citet{Green2015}.}
  \label{sfield}
\end{figure*}

\begin{figure*}
  \centering
  \includegraphics[width=0.88\textwidth]{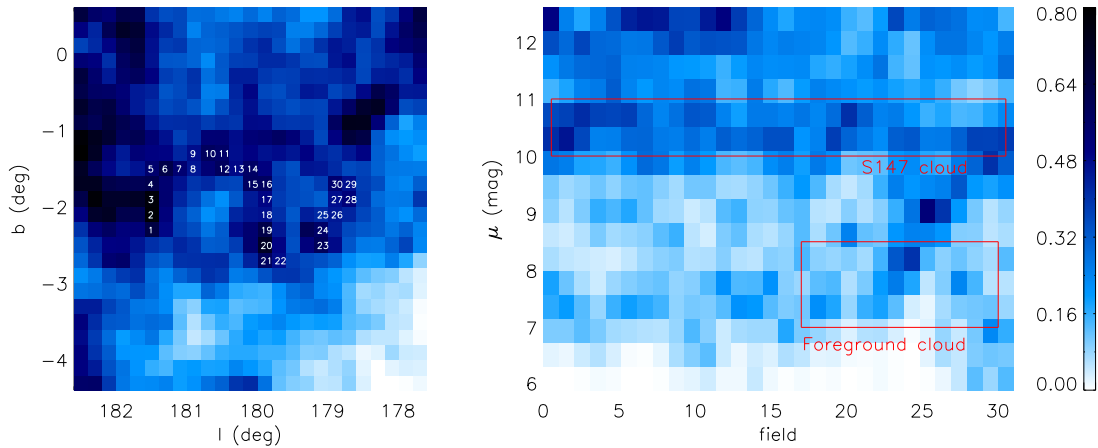}
  \caption{Distributions of dust extinction as a function of distance 
for subfields of the S147 dust cloud.  The left panel shows the distribution of dust in
distance bin from 1000 to 1585\,pc. The map has a resolution of 0.2\degr. 
The selected pixels belonging to the S147 dust cloud are 
used for a precise analysis of distance to the cloud, are labelled with a
unique number from 1 to 30. Each pixel is of size $0.2\degr\ \times 0.2\degr$.
The right panel shows the amounts of extinction $A_r$ for the 30 pixels 
(subfields) in different distance bins. 
The X-axis denotes the unique number of the pixel.
The red squares mark roughly the boundaries of pixels belonging to the 
foreground cloud and the S147 dust cloud,
respectively.}
  \label{slice}
\end{figure*}

\begin{figure}
  \centering
  \includegraphics[width=0.48\textwidth]{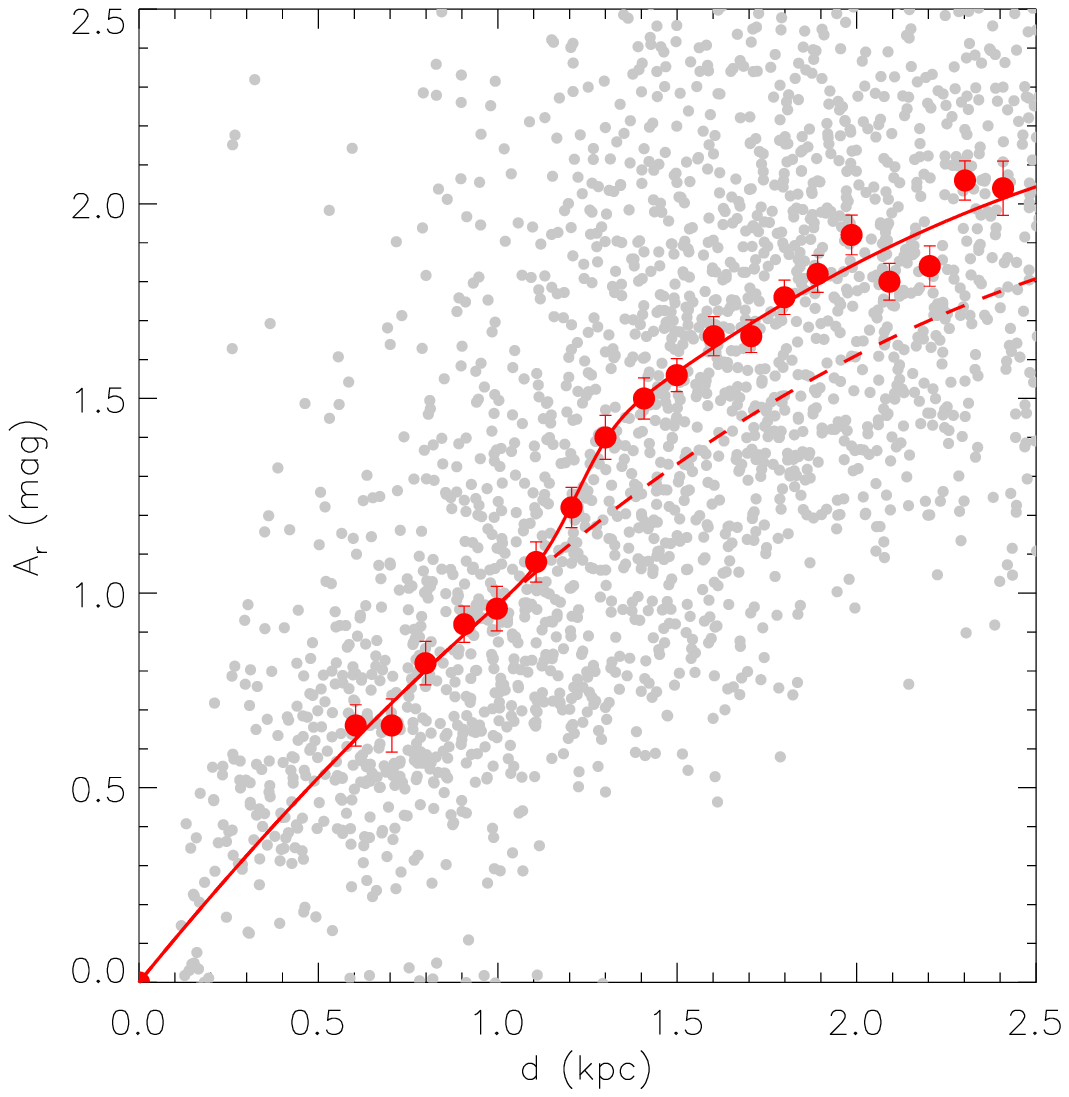}
  \caption{Extinction values plotted against distances for stars in the 
 selected subfields No.\,1 to 22 (Fig.~\ref{slice}). 
 The red circles are median values in the individual distance bins starting from 0.5\,kpc. 
 The errorbars are the standard errors. The red solid line
 is the best-fit extinction law based on the median values. 
 The red dashed line is the modeled extinction law without 
 the contribution of the S147 dust cloud.}
  \label{dis}
\end{figure}

From the above 3D dust maps, it is easy to infer that the S147 dust cloud 
has a distance between 1.0 and 1.6\,kpc.
The accuracy of this inferred distance is limited by uncertainties of distances
of individual stars.
The uncertainties of distances of individual stars, 
which are estimated to be 20\,per\,cent, arise from 
uncertainties in measured colours of the stars, in constructed library of stellar 
loci, as well as from uncertainties in the distribution of stellar ages and metallicities.
Our distance resolution is also limited by 
pixel  size of $6^\prime \times 6^\prime$ used to fit the 3D dust distribution profiles.
A better approach to estimate the distance of the S147 dust cloud  
is to use larger pixels with a single, common dust distribution  profile,  
similar to the technique adopted by  \citet{Schlafly2014} 
for the estimation of distances to MCs. However, 
one should note that when \citet{Schlafly2014} used this 
technique to estimate distances to MCs, they 
assumed that the dust extinction is dominated by a single cloud. This is not true in our case.
In Fig.~\ref{sfield} we plot the distributions of $r$-band extinction 
values as a function of distance for stars
in five representative pixels (subfields)  toward S147.  
Also overplotted in the Figure are the best-fit extinction and distance 
relations for those lines of sight from \citet{Green2015}. 
\citet{Green2015} obtain their 3D dust distribution based on the
Pan-STARRS 1 (PS1; \citealt{Kaiser2002}) and 2MASS photometry. The map has a 
resolution of about 5$^\prime$ in the area of interest here. 
For each line of sight, their best fit reddening and distance relation
nearest to the pixel is used. Their reddening values
are converted to $r$-band extinction assuming $A_r = 2.32 E(B-V)$ \citep{Yuan2013}. 
In general, our deduced extinction profiles are in good agreement 
with those of \citet{Green2015}, except that at large distances ($d$ $>$ 1.5\,kpc), 
\citet{Green2015} obtain higher extinction than ours. This is mainly due to the higher angular resolution 
of their map and deeper photometry used. Extinction values and distances 
available from LSS-GAC DR2 are also plotted in
the Figure. Again the extinction profiles derived here match the spectroscopic results well. 
The vertical shadows overplotted in Fig.~\ref{sfield} delineate respectively the distance ranges 
of the foreground cloud (300$<d<$600\,kpc) and of the S147 cloud (1000$<d<$1500\,kpc). 
A precise distance estimate for the foreground cloud 
is difficult given that there is not enough stars in front of 
the dust in our sample, due to the bright saturation limit of XSTPS-GAC and the 
high angular resolution of the extinction map adopted here. 
In Fig.~\ref{sfield}, the best fit extinction profiles
(red lines in the individual panels) of the selected lines of sight No.~1, 2 and 3 have  
a ``jump'' in extinction value at a distance between 1000 and 1500\,pc. It is not surprising to see some  
variations in ``jump'' distances  for different pixels,  
given a pixel size of only 0.1\degr\ $\times$ 0.1\degr\ 
and a bin size of distance module of 0.5\,mag (about 300\,pc at 1000\,pc). 
For line of sight No.~4, only a ``jump'' in extinction attributed to the foreground cloud 
is seen. For line of sight No.~5,  two  ``jumps'' are visible, produced by   
the foreground and the S147 cloud, respectively. 

In Fig.~\ref{slice} we investigate possible variations in 
distance of different portion of the S147 dust cloud.
For this purpose, we have repeated the 3D dust mapping 
procedure with a lower resolution of 0.2\degr. The narrow filamentary features  
and the individual small clumps become poorly resolved at this resolution, 
but the resultant map of dust distribution is 
qualitatively similar. We then select pixels belonging to the S147 dust cloud 
and investigate the amounts of dust extinction of those pixels in different distances 
bins. Thirty pixels are selected at 
Galactic longitudes between 178\degr\ and 182\degr\, running   
through the entire cloud and labelled by a number from 1 to 30. 
The amounts of extinction of these pixels are shown in Fig.~\ref{slice}. 
The maps have been smoothed 
with a Gaussian kernel of 0.5\degr\ FWHM for better illustration. 
The S147 dust cloud appears prominently in Fig.~\ref{slice} 
as a dark blue snake at distance modulus $10 < \mu < 11$\,mag, 
i.e. of distances $1.0 < d < 1.5$\,kpc, for all the pixels.
Contamination from the foreground cloud is also apparent in pixels 
of number~23 to 30.

Finally, we select all stars in the fields labelled from No.~1 to 22. 
The above analysis shows that 
the extinction observed in those regions are mainly contributed by a single cloud, i.e. the S147 cloud. 
The variations of extinction as a function of distance of all stars in these pixels
are plotted in Fig.~\ref{dis}. The scatter of extinction values at a given distance of these stars is large. 
And this may be attributed to the fact that the stars cover 
a large area (nearly 1\,deg$^2$). 
Yet via  the binned median extinction values, one  can clearly see  the 
extinction ``jump'' produced by the S147 dust cloud at a distance  between 1100 and 1300\,pc. 
A simple extinction model is applied to 
estimate the distance and amplitude of the  ``jump''. 
We assume a extinction-distance relation as following,
\begin{equation}
  A_r(d) = A_r^0 (d) + A_r^1 (d),
\end{equation}
where $A_r^1(d)$ is the contribution of extinction from the S147 cloud, i.e. the 
``jump'' component. We assume that this component can be simply described by a Gaussian 
function. Thus the integrated extinction contributed  the S147 cloud to a distance $d$ is given by, 
\begin{equation}
 A_r^{1}(d)= \frac{\delta A_r}{2} {\rm erf}(\frac{d-d_0}{ \sqrt{2}* \delta d}),
\end{equation}
where $\delta A_r$ is the extinction ``jump'' amplitude, $\delta d$ is the width of the cloud and
$d_0$ is the distance of peak extinction (i.e. the centre of the S147 dust cloud). 
In Eq.~(5), $A_r^0$ represents the smooth extinction component 
contributed by the general interstellar medium and is assume to be
a 2-order polynomial \citep{Chen1998},
\begin{equation}
 A_r^0(d)=ad+bd^2,
\end{equation}
where $a$ and $b$ are polynomial coefficients. The binned average extinction values for 
the distance range $0.6 < d < 2.5$\,kpc are fitted with the 
extinction model described above using a maximum likelihood method. The
best-fit result is overplotted in Fig.~\ref{dis}. The fit yields  
$\delta A_r = 0.24\pm 0.28$\,mag, $\delta d$=81$\pm 299$\,pc and $d_0$=1223$\pm 206$\,pc. 
This distance to the S147 dust cloud yielded by our analysis, $d=$1.22$\pm$0.21\,kpc, 
is consistent with the most recent distance estimates of S147, 
either based on the plausible association of S147 with  
pulsar PSR J0538 + 2817 \citep{Kramer2003,Ng2007,Chatterjee2009} 
or its pre-supernova binary companion HD\,37424 (\citealt{Dincel2015}, see Table~1). 
The latest estimate of distance to the pulsar is 1.30$^{+0.22}_{-0.16}$\,kpc, based on  
the parallax measurement of the pulsar \citep{Chatterjee2009}, 
while that based on its pre-supernova binary companion is
1.333$^{+0.103}_{-0.112}$\,kpc \citep{Dincel2015}. Both results are in good agreement with
our estimate of distance to the S147 dust cloud. 
The shorter distance, $d<$ 0.88 kpc, obtained via an absorption line
 analysis of the B1e star HD\,36665 (which is assumed to be formed out of the  gas associated 
with S147; \citealt{Phillips1981, Sallmen2004}), may be caused by the contamination of the 
foreground cloud. Gas detected in the absorption line 
spectrum of HD\,36665 might in fact originated from 
the foreground cloud, which is estimated to have a distance $d<631$\,pc.

From the above model, the integrated dust extinction toward the S147 dust cloud 
is $A_r = 1.26$\,mag  ($A_V = 1.48$\,mag). 
We note however there are significant variations in the total 
extinction across the surface of the cloud. 
Former extinction estimates of S147, $A_V =1.28$\,mag by \citet{Dincel2015}  
and $A_V = 0.7$\,mag by \citet{Fesen1985}, 
are both based on observation of a small area of the cloud. 
The above best-fit model also yields $\delta A_r = 0.24\pm 0.28$\,mag,
representing the average extinction contributed by the cloud. 
For some parts of the cloud, such as the three 
dense clumps  mentioned above, the total extinction could be as large as $\delta A_r \sim 1$\,mag.

\subsection{Mass}

The mass of the S147 dust cloud is dominated by atomic and molecular hydrogen. We 
can make a rough estimate of the mass of the cloud by converting the 
observed extinction $A_V$ to mass.  Following 
\citet{Lombardi2011} and \citet{Schlafly2015}, the total mass $M$ of the cloud 
is given by, 
\begin{equation}
  M=\frac{d^2\mu }{DGR}\int{A_V(\Omega) d\Omega}
\end{equation}
where $d$ is the distance to the cloud, $\mu$ the mean molecular weight,
$A_V({\Omega})$ the optical extinction $A_V$ in solid angle ${\Omega}$,
and $DGR$ is the dust-to-gas ratio, 
\begin{equation}
DGR = A_V/N({\rm H}),
\end{equation}
where $N({\rm H})=N({\rm H~{\scriptstyle I}})+2N({\rm H_2})$ is the total hydrogen column density. 
We convert our $r$-band extinction to $V$-band 
using the extinction law of \citet{Yuan2013}, which yields
$A_V=1.17A_r$. We adopt $\mu = 1.37$ from \citet{Lombardi2011} and 
$DGR=4.15 \times 10^{-22}~\mathrm{mag}~\mathrm{cm}^{2}$ from \citet{Chen2015}. The
boundary of the S147 cloud is defined as $178.6\degr \le l  \le 182.0\degr;$ and 
$ -3.2\degr \le b \le 0\degr$.
We assume that only dust within distance range 1000 $< d <$ 1585\,pc is 
associated with S147, 
and the dust is distributed  as a thin screen at a distance of $d$=1.2\,kpc.

This leads to a total mass of the S147 cloud of 176,315$M_\odot$. 
The estimate is certainly quite rough, considering the many simplifications involved.
The accuracy is mainly limited to the uncertainties of adopted distance.
In the analysis we have assumed a distance of 1.2\,kpc. 
The uncertainty in the distance could easy make the estimated mass to change by a factor of two.
Moreover, due to the uncertainty in distance estimate, 
some of the dust actually belonging to the foreground or 
background medium could be wrongly attributed  
to the S147 dust cloud, or vice versa. 
In the current calculation, we have accepted all  dust in a large range of estimated distances, 
1000 $< d <$ 1585\,pc, as belong to the  S147 cloud. 
Considering that in the case of S147, there is significant dust 
presence both in the foreground and background, some contamination 
from the foreground and background material is somehow inevitable. Thus the total mass 
obtained above could be overestimated.
The choice of  $DGR$ value may also lead to additional uncertainties. The most cited value,  
$5.3 \times 10^{-22}~\mathrm{mag}~\mathrm{cm}^{2}$, derived by \citet{Bohlin1978} is slightly 
larger than the adopted value here. \citet{Chen2015} find that 
the values of $DGR$ for the individual clouds in the Galactic anticentre can vary by a factor of two. 
In spite of all of these uncertainties, the analysis here still place some useful constraints on the mass
of the S147 dust cloud\footnote{Note that the S147 dust cloud is defined as part of a large dust cloud. To estimate the mass of the entire dust cloud, data over a larger field are needed. The analysis is however beyond the scope of the current work.}. The estimated mass,  approximately 10$^5$\,$M_\odot$, 
is however of the same order of magnitude of those estimates for
the well known molecular clouds like Orion, Monoceros R2, Rosette and the Canis Major \citep{Lombardi2011}.

\section{Morphology of the S147 dust cloud }

\begin{figure*}
  \centering
  \includegraphics[width=0.98\textwidth]{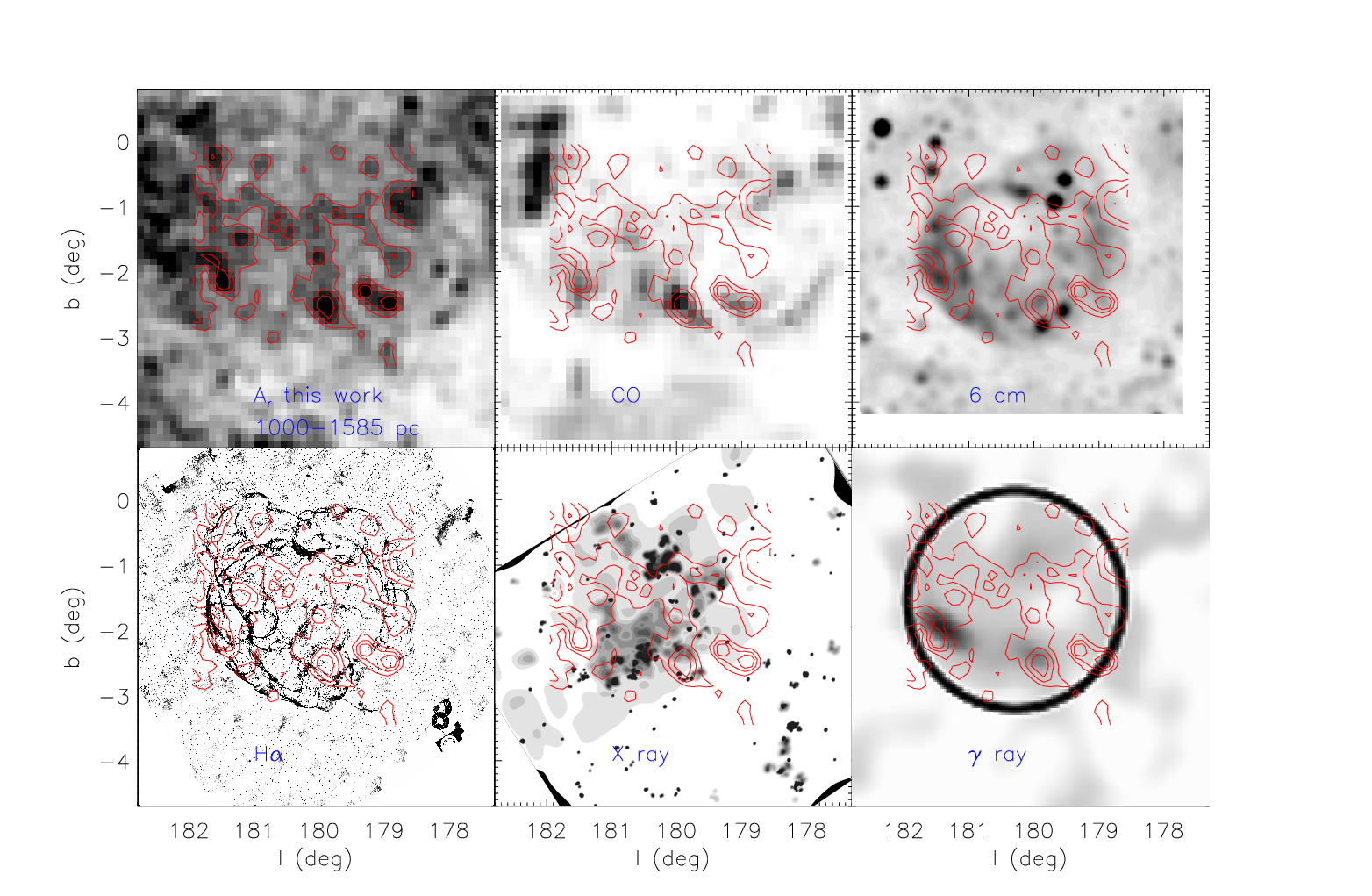}
  \caption{The extinction distribution of S147 dust cloud is compared to 
millimetre  CO, radio 6\,cm continuum, optical \Ha, X-ray, 
and gamma-ray emission maps of the region of interest.  The data used are: CO from \citet{Dame2001},
radio 6\,cm from \citet{Xiao2008}, \Ha\  from \citet{Drew2005},
X-ray from ROSAT \citep{Voges1999} for the 
energy range 0.4 -- 0.9\,keV, and gamma-ray from 
\citet{Katsuta2012}. The black circle in the last panel marks S147. 
The red contours in each panel shows the dust extinction 
distribution of the S147 dust cloud.}
  \label{others}
\end{figure*}

In this section, we compare the morphology of the newly 
identified S147 dust cloud to maps of S147 viewed in
other frequencies, in order to investigate the plausible correlation 
between the dust and other tracers, such as cool or hot gas, 
in the region of interest. The comparisons  is presented  in
Fig.~\ref{others}, where we compare the dust extinction distribution  
to the S147 images in millimetre  from the CO, 6\,cm
radio continuum , optical {\rm \Ha}, X-ray and 
gamma-ray emission, respectively.

\subsection{CO emission}

\begin{figure}
  \centering
  \includegraphics[width=0.48\textwidth]{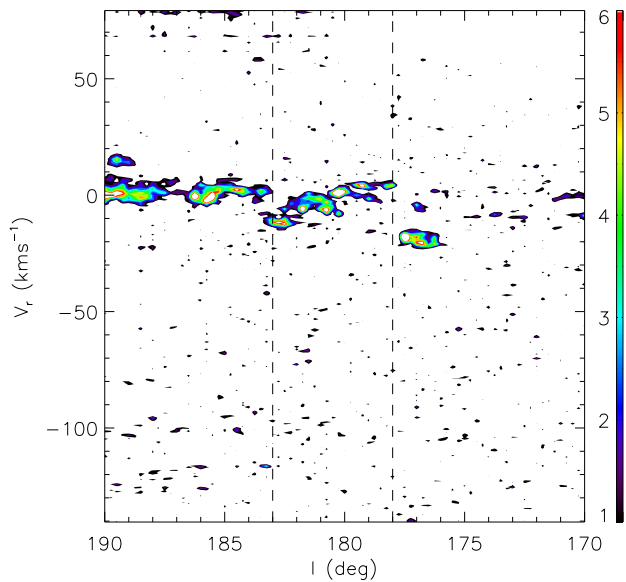}
  \caption{Galactic longitude and velocity diagram of CO gas 
  from \citet{Dame2001} for the region of $-3\degr\ < b < -1\degr$.  The 
  vertical dashed lines  mark  the longitude range of the S147 dust cloud.}
  \label{covr}
\end{figure}

The second panel of Fig.~\ref{others} shows the CO emission map from 
\citet{Dame2001} in the region of interest.  
\citet{Dame2001} combine large-scale CO surveys of the Galactic plane and of all large local clouds at 
higher latitudes into a composite map of the entire Milky Way at an angular resolution ranging 
from 9$^\prime$ to 18$^\prime$. 
The panel shows the total CO column density in the area, with contours of the S147 dust cloud overplotted.  
Most prominent features including the three dense clumps seen 
in the dust map are clearly detected in CO emission.  
On the other hand, regions of less dust extinction have no detectable CO,
presumably as a result of the low column density of dust
grains in those regions. CO molecules have been be largely photodissociated
due to insufficient shielding.

\citet{Huang1986} carried out a CO survey toward all SNRs from 
Galactic longitude 70\degr\ to 210\degr.
They find two  large MCs, coincident in position with 
the three dense dust clumps of S147 found here. 
Recently, \citet{Jeong2012} carried out a CO survey of SNRs between longitudes 70\degr\ -- 190\degr\
using the Seoul Radio Astronomy Observatory (SRAO) 6-m radio telescope. 
They have also found some molecular 
clouds, at velocities ranging from $\sim$  $-$14 to $+$5 \kms, 
that overlap in position with the dense dust regions of 
S147. Both studies point out that the MCs detected do not show features correlated with the 
S147 SNR. This may be partly caused  by the lack of distance  information. 
Also, one should bear in mind that the CO ($J~=$~1 -- 0) may
not be a good tracer for the diffused  gas of  low dust column densities. 
The 3D dust mapping technique employed here not only allows us to 
measure a distance to the S147 dust cloud, but also uncover the diffuse parts of the cloud. 
We believe that 3D dust extinction could be a better tracer to trace the SNR 
associated MCs than the CO ($J~=$~1 -- 0)  emission. 

Fig.~\ref{covr} shows the observed CO \citep{Dame2001} distribution in the direction of Galactic anticentre
within $-3\degr < b < -1\degr$ in the $l$ and $V_r$ space. 
The black lines in the Figure enclose the feature
of the S147 dust cloud. The velocity of the cloud spans from $-$10 to 5 \kms, in consistent 
with the observation of \citet{Huang1986} and \citet{Jeong2012}. The peak velocities in regions of smaller 
Galactic longitudes are slightly higher than those of higher Galactic longitudes. In particular, 
for the clouds of $l < $180\degr, the radial velocities of CO 
emitting gas are positive, while of $l > $180\degr\ are negative. 
This could be explained by the expansion of the cloud. Consequently, the variations seen in velocity  
does not clearly correlate with the variations observed in distance. In principle,  
observation of neutral hydrogen atoms should be sensitive 
enough to trace the gas associated with the dust cloud,
especially for regions lacking strong CO emission. 
Unfortunately, in the region of S147, 
any \HI\ emission associated with the cloud is lost in strong background of  
the Galactic \HI\ disk. It could be a challenge to isolate the clumps and filaments correlated with the 
dust cloud found in the current work. 

\subsection{6\,cm radio emission}

The third panel of Fig.~\ref{others} shows the radio 6\,cm map of S147 \citep{Xiao2008}.
The map gives the total intensity of radio continuum of S147 at 
6\,cm (4800\,MHz), obtained with the Urumqi 25-m 
telescope between January 2005 and February 2006. 
It has an angular resolution of 9.51$^{\prime}$. Again
overplotted in the panel is the $r$-band extinction 
contours of the S147 dust cloud. The 6\,cm map shows a spherical SNR 
shell with a hollow region at the centre,  quite different from the morphology of dust distribution. 
Although one does not expect 
any  good correlation between the dust extinction and the 
emission of ionised gas as traced by the 6\,cm observation, one does see
some level of correlation between them in the Figure.
At  the locations of two out of three dense clumps of the S147 dust cloud 
(one at $l ~\sim$ 181.3\degr, $b~\sim~-$2.2\degr\, 
and another at  $l ~\sim$ 179.9\degr, $b~\sim~-$2.7\degr), the 
6\,cm emission is also strong. 
%On the other hand, all the dense clumps
%seen in dust extinction fall either on or outside the radio shell of S147. Inside the radio shell, 
%one sees only some diffuse dust of relatively low column densities.  
%Presumably, all dense dust clumps inside the radio shell have 
%been destroyed or swept out by the SN explosion. 

\subsection{\Ha\ emission}

The fourth panel of Fig.~\ref{others} shows the \Ha\ image of S147 from the 
INT Photometric \Ha\ Survey of the Northern 
Galactic Plane (IPHAS; \citealt{Drew2005}). The \Ha\ image shows beautiful ring-like filaments
and has similar features as the 6\,cm map.
Like the 6\,cm map, the \Ha\ image shows complicated 
spatial correlations with the dust cloud we found in this paper.
A detailed kinematic analysis of the
optical filaments and shells will be presented in a separate work 
(Ren et al., in preparation).

\subsection{X-ray emission}

Supernova remnants often emit strong X-ray emission. 
\citet{Sauvageot1990} report that 
no X-ray emission is observed from S147 with the X-ray satellite EXOSAT.
\citet{Sun1996} claim that X-ray emission from S147 has been unambiguously detected
in the ROSAT all-sky survey. In the fifth panel of Fig.~\ref{others} we show
the X-ray image in the energy range of 0.4 -- 0.9\,keV from the ROSAT all sky survey \citep{Voges1999}
for the region of interest. The X-ray image has a patchy appearance without a clear
shell, with some similarity with the 6\,cm image. 
There is anti-correlation between the X-ray emission and the dust extinction. 
In regions where the X-ray emission is strong, 
the extinction is relative low, and vice versa. 
This is mainly due to the fact that
the dust cloud may have absorbed all the low-energy X-rays produced behind it.

\subsection{Gamma-ray emission}

Finally we show in Fig.~\ref{others} the gamma-ray image of S147 \citep{Katsuta2012}. 
\citet{Katsuta2012} have analysed the gamma-ray data obtained with
the Large Area Telescope on board the Fermi Gamma-ray Space Telescope
\citep{Atwood2009} in the region toward S147. 
They create a background-subtracted count map of the region above 1\,GeV (showed in the right panel
of Fig.~1 of \citealt{Katsuta2012}), which is reproduced here. There is 
spatially extended gamma-ray emission of energy range 0.2--10\,GeV in the region. 
Two of the prominent dense dust clumps centred at
($l$, $b$) = (181.3\degr, $-2.2$\degr) and ($l$, $b$) = (179.9\degr, $-$0.4\degr) spatially 
coincident with the gamma-ray emission.

The gamma-ray emission surrounding the SNR is expected to be primarily 
emitted via the pion decay produced by the interaction between CR protons and dense 
medium (p-p interaction), while the leptonic origin is unlikely for such an old 
SNR \citep{Aharonian1994, Ackermann2013}. The morphology correspondence 
between hadronic gamma-ray emission and MC has been found in a 
number of evolved SNR (e.g., W44, IC443; \citealt{Seta1998}).
The comparisons of gamma-ray emission and dust 
distribution in the direction of S147 as presented above shows
a possible correlation between them, suggesting probable interactions 
between S147 and its surrounding MCs \citep{Katsuta2012}. Finally,
 \citet{Katsuta2012} model the gamma-ray spectrum of S147
and obtain a gas density between 100 -- 500\,cm$^{-3}$, 
corresponding to a dust cloud of extinction $A_V$ $\sim$ 0 -- 1\,mag. This is consistent with the 
estimate of extinction of the S147 cloud obtained in this paper.

\section{Conclusion}

In this work, we present a 3D map of dust extinction toward the SNR S147.  
When integrated to a distance of about 2.5\,kpc, our map traces similar dust features  
as revealed by the {\it Planck} data. We have estimated distances
to the several dust  clouds detected in the map. 
The work reveals a large dust cloud at a distance between 1 $<~d~<$ 1.5\,kpc.
Parts of the cloud are possibly associated with the SNR,
and are termed as the ``S147 dust cloud.'' 
The cloud is consistent of several dense clumps of high extinction, 
which locate on the radio shell of S147,
as well as some diffuse parts of low extinction.
We estimate the distance to the cloud using a simple extinction
model. This yields a distance of 1.22$\pm$0.21\,kpc, 
in consistent with the distance estimated for the SNR. 
We have also obtained a rough estimate of the mass of the S147 dust cloud, which is
about 10$^5$\,$\rm M_\odot$. 

We compare the dust extinction distribution of the S147 dust cloud 
with the images of S147 in other frequencies, including those in the CO, radio 6\,cm, \Ha,  
X-ray and gamma-ray emission. The dust distribution is not surprisingly 
correlated with the CO emission. More interestingly, 
the dust distribution is spatially correlated with the 
gamma-ray emission in energy range 0.2 -- 10\,GeV,
indicating that the dust cloud is possibly associated with the SNR
and has complicated interactions with the SNR. 

We conclude that comparing to the CO ($J$=1--0) emission, 3D dust mapping
could be a viable technique to search for MCs associated with SNRs, 
and to study the SNR-MC interactions. In a future work, we 
will present a systematic survey of MCs associated with SNRs 
in the outer disk of the Galaxy (Yu et al., in preparation). Utilizing 
tens of thousands of stellar spectra collected as parts of LSS-GAC survey
in the vicinity of S147,  we are able to study the extinction law,
referring the dust size distribution, as well as to carry out a demographic 
study of the DIBs of the S147 dust cloud. 
A related work on these lines will be presented in near future (Chen et al., in preparation).

\section*{Acknowledgements}

We thank our anonymous referee for helpful comments 
that improved the quality of this paper. B.Q.C. thanks
Professors Biwei Jiang, Jian Gao, Jun Fan, Yang Chen and Ping Zhou for very useful comments.
This work is partially supported by National Key Basic Research Program of China
2014CB845700 and  China Postdoctoral Science Foundation 2016M590014. 
The LAMOST FELLOWSHIP is supported by Special Funding for Advanced Users,
budgeted and administrated by Center for Astronomical Mega-Science, Chinese
Academy of Sciences (CAMS). 

This work has made use of data products from the Guoshoujing Telescope (the
Large Sky Area Multi-Object Fibre Spectroscopic Telescope, LAMOST). LAMOST
is a National Major Scientific Project built by the Chinese Academy of
Sciences. Funding for the project has been provided by the National
Development and Reform Commission. LAMOST is operated and managed by the
National Astronomical Observatories, Chinese Academy of Sciences.

\bibliographystyle{mn2e}
\bibliography{extinsnr}

\begin{thebibliography}{65}
\expandafter\ifx\csname natexlab\endcsname\relax\def\natexlab#1{#1}\fi

\bibitem[{{Ackermann} {et~al.}(2013){Ackermann}, {Ajello}, {Allafort},
  {Baldini}, {Ballet}, {Barbiellini}, {Baring}, {Bastieri}, {Bechtol},
  {Bellazzini}, {Blandford}, {Bloom}, {Bonamente}, {Borgland}, {Bottacini},
  {Brandt}, {Bregeon}, {Brigida}, {Bruel}, {Buehler}, {Busetto}, {Buson},
  {Caliandro}, {Cameron}, {Caraveo}, {Casandjian}, {Cecchi}, {{\c C}elik},
  {Charles}, {Chaty}, {Chaves}, {Chekhtman}, {Cheung}, {Chiang}, {Chiaro},
  {Cillis}, {Ciprini}, {Claus}, {Cohen-Tanugi}, {Cominsky}, {Conrad}, {Corbel},
  {Cutini}, {D'Ammando}, {de Angelis}, {de Palma}, {Dermer}, {do Couto e
  Silva}, {Drell}, {Drlica-Wagner}, {Falletti}, {Favuzzi}, {Ferrara},
  {Franckowiak}, {Fukazawa}, {Funk}, {Fusco}, {Gargano}, {Germani},
  {Giglietto}, {Giommi}, {Giordano}, {Giroletti}, {Glanzman}, {Godfrey},
  {Grenier}, {Grondin}, {Grove}, {Guiriec}, {Hadasch}, {Hanabata}, {Harding},
  {Hayashida}, {Hayashi}, {Hays}, {Hewitt}, {Hill}, {Hughes}, {Jackson},
  {Jogler}, {J{\'o}hannesson}, {Johnson}, {Kamae}, {Kataoka}, {Katsuta},
  {Kn{\"o}dlseder}, {Kuss}, {Lande}, {Larsson}, {Latronico}, {Lemoine-Goumard},
  {Longo}, {Loparco}, {Lovellette}, {Lubrano}, {Madejski}, {Massaro}, {Mayer},
  {Mazziotta}, {McEnery}, {Mehault}, {Michelson}, {Mignani}, {Mitthumsiri},
  {Mizuno}, {Moiseev}, {Monzani}, {Morselli}, {Moskalenko}, {Murgia},
  {Nakamori}, {Nemmen}, {Nuss}, {Ohno}, {Ohsugi}, {Omodei}, {Orienti},
  {Orlando}, {Ormes}, {Paneque}, {Perkins}, {Pesce-Rollins}, {Piron}, {Pivato},
  {Rain{\`o}}, {Rando}, {Razzano}, {Razzaque}, {Reimer}, {Reimer}, {Ritz},
  {Romoli}, {S{\'a}nchez-Conde}, {Schulz}, {Sgr{\`o}}, {Simeon}, {Siskind},
  {Smith}, {Spandre}, {Spinelli}, {Stecker}, {Strong}, {Suson}, {Tajima},
  {Takahashi}, {Takahashi}, {Tanaka}, {Thayer}, {Thayer}, {Thompson},
  {Thorsett}, {Tibaldo}, {Tibolla}, {Tinivella}, {Troja}, {Uchiyama}, {Usher},
  {Vandenbroucke}, {Vasileiou}, {Vianello}, {Vitale}, {Waite}, {Werner},
  {Winer}, {Wood}, {Wood}, {Yamazaki}, {Yang}, \& {Zimmer}}]{Ackermann2013}
{Ackermann}, M., {et~al.} 2013, Science, 339, 807

\bibitem[{{Aharonian} {et~al.}(1994){Aharonian}, {Drury}, \&
  {Voelk}}]{Aharonian1994}
{Aharonian}, F.~A., {Drury}, L.~O., \& {Voelk}, H.~J. 1994, \aap, 285, 645

\bibitem[{{Anderson} {et~al.}(1996){Anderson}, {Cadwell}, {Jacoby},
  {Wolszczan}, {Foster}, \& {Kramer}}]{Anderson1996}
{Anderson}, S.~B., {Cadwell}, B.~J., {Jacoby}, B.~A., {Wolszczan}, A.,
  {Foster}, R.~S., \& {Kramer}, M. 1996, \apjl, 468, L55

\bibitem[{{Atwood} {et~al.}(2009){Atwood}, {Abdo}, {Ackermann}, {Althouse},
  {Anderson}, {Axelsson}, {Baldini}, {Ballet}, {Band}, {Barbiellini}, \&
  et~al.}]{Atwood2009}
{Atwood}, W.~B., {et~al.} 2009, \apj, 697, 1071

\bibitem[{{Berry} {et~al.}(2012){Berry}, {Ivezi{\'c}}, {Sesar}, {Juri{\'c}},
  {Schlafly}, {Bellovary}, {Finkbeiner}, {Vrbanec}, {Beers}, {Brooks},
  {Schneider}, {Gibson}, {Kimball}, {Jones}, {Yoachim}, {Krughoff}, {Connolly},
  {Loebman}, {Bond}, {Schlegel}, {Dalcanton}, {Yanny}, {Majewski}, {Knapp},
  {Gunn}, {Allyn Smith}, {Fukugita}, {Kent}, {Barentine}, {Krzesinski}, \&
  {Long}}]{Berry2012}
{Berry}, M., {et~al.} 2012, \apj, 757, 166

\bibitem[{{Bohlin} {et~al.}(1978){Bohlin}, {Savage}, \& {Drake}}]{Bohlin1978}
{Bohlin}, R.~C., {Savage}, B.~D., \& {Drake}, J.~F. 1978, \apj, 224, 132

\bibitem[{{Chatterjee} {et~al.}(2009){Chatterjee}, {Brisken}, {Vlemmings},
  {Goss}, {Lazio}, {Cordes}, {Thorsett}, {Fomalont}, {Lyne}, \&
  {Kramer}}]{Chatterjee2009}
{Chatterjee}, S., {et~al.} 2009, \apj, 698, 250

\bibitem[{{Chen} {et~al.}(1998){Chen}, {Vergely}, {Valette}, \&
  {Carraro}}]{Chen1998}
{Chen}, B., {Vergely}, J.~L., {Valette}, B., \& {Carraro}, G. 1998, \aap, 336,
  137

\bibitem[{{Chen} {et~al.}(2015){Chen}, {Liu}, {Yuan}, {Huang}, \&
  {Xiang}}]{Chen2015}
{Chen}, B.-Q., {Liu}, X.-W., {Yuan}, H.-B., {Huang}, Y., \& {Xiang}, M.-S.
  2015, \mnras, 448, 2187

\bibitem[{{Chen} {et~al.}(2014{\natexlab{a}}){Chen}, {Liu}, {Yuan}, {Zhang},
  {Schultheis}, {Jiang}, {Huang}, {Xiang}, {Zhao}, {Yao}, \& {Lu}}]{Chen2014}
{Chen}, B.-Q., {et~al.} 2014{\natexlab{a}}, \mnras, 443, 1192

\bibitem[{{Chen} {et~al.}(2013){Chen}, {Schultheis}, {Jiang}, {Gonzalez},
  {Robin}, {Rejkuba}, \& {Minniti}}]{Chen2013}
{Chen}, B.~Q., {Schultheis}, M., {Jiang}, B.~W., {Gonzalez}, O.~A., {Robin},
  A.~C., {Rejkuba}, M., \& {Minniti}, D. 2013, \aap, 550, A42

\bibitem[{{Chen} {et~al.}(2014{\natexlab{b}}){Chen}, {Jiang}, {Zhou}, {Su},
  {Zhou}, {Li}, \& {Zhang}}]{Chen2014b}
{Chen}, Y., {Jiang}, B., {Zhou}, P., {Su}, Y., {Zhou}, X., {Li}, H., \&
  {Zhang}, X. 2014{\natexlab{b}}, in IAU Symposium, Vol. 296, Supernova
  Environmental Impacts, ed. A.~{Ray} \& R.~A. {McCray}, 170--177

\bibitem[{{Clark} \& {Caswell}(1976)}]{Clark1976}
{Clark}, D.~H. \& {Caswell}, J.~L. 1976, \mnras, 174, 267

\bibitem[{{Dame} {et~al.}(2001){Dame}, {Hartmann}, \& {Thaddeus}}]{Dame2001}
{Dame}, T.~M., {Hartmann}, D., \& {Thaddeus}, P. 2001, \apj, 547, 792

\bibitem[{{Din{\c c}el} {et~al.}(2015){Din{\c c}el}, {Neuh{\"a}user}, {Yerli},
  {Ankay}, {Tetzlaff}, {Torres}, \& {Mugrauer}}]{Dincel2015}
{Din{\c c}el}, B., {Neuh{\"a}user}, R., {Yerli}, S.~K., {Ankay}, A.,
  {Tetzlaff}, N., {Torres}, G., \& {Mugrauer}, M. 2015, \mnras, 448, 3196

\bibitem[{{Drew} {et~al.}(2005){Drew}, {Greimel}, {Irwin}, {Aungwerojwit},
  {Barlow}, {Corradi}, {Drake}, {G{\"a}nsicke}, {Groot}, {Hales}, {Hopewell},
  {Irwin}, {Knigge}, {Leisy}, {Lennon}, {Mampaso}, {Masheder}, {Matsuura},
  {Morales-Rueda}, {Morris}, {Parker}, {Phillipps}, {Rodriguez-Gil}, {Roelofs},
  {Skillen}, {Sokoloski}, {Steeghs}, {Unruh}, {Viironen}, {Vink}, {Walton},
  {Witham}, {Wright}, {Zijlstra}, \& {Zurita}}]{Drew2005}
{Drew}, J.~E., {et~al.} 2005, \mnras, 362, 753

\bibitem[{{Fesen} {et~al.}(1985){Fesen}, {Blair}, \& {Kirshner}}]{Fesen1985}
{Fesen}, R.~A., {Blair}, W.~P., \& {Kirshner}, R.~P. 1985, \apj, 292, 29

\bibitem[{{Frail} {et~al.}(1996){Frail}, {Goss}, {Reynoso}, {Giacani}, {Green},
  \& {Otrupcek}}]{Frail1996}
{Frail}, D.~A., {Goss}, W.~M., {Reynoso}, E.~M., {Giacani}, E.~B., {Green},
  A.~J., \& {Otrupcek}, R. 1996, \aj, 111, 1651

\bibitem[{{Gaia Collaboration} {et~al.}(2016){Gaia Collaboration}, {Brown},
  {Vallenari}, {Prusti}, {de Bruijne}, {Mignard}, {Drimmel}, {Babusiaux},
  {Bailer-Jones}, {Bastian}, \& et~al.}]{Gaia2016}
{Gaia Collaboration}, {et~al.} 2016, \aap, 595, A2

\bibitem[{{Goodman} {et~al.}(2009){Goodman}, {Pineda}, \&
  {Schnee}}]{Goodman2009}
{Goodman}, A.~A., {Pineda}, J.~E., \& {Schnee}, S.~L. 2009, \apj, 692, 91

\bibitem[{{Green} {et~al.}(1997){Green}, {Frail}, {Goss}, \&
  {Otrupcek}}]{Green1997}
{Green}, A.~J., {Frail}, D.~A., {Goss}, W.~M., \& {Otrupcek}, R. 1997, \aj,
  114, 2058

\bibitem[{{Green} {et~al.}(2014){Green}, {Schlafly}, {Finkbeiner}, {Juri{\'c}},
  {Rix}, {Burgett}, {Chambers}, {Draper}, {Flewelling}, {Kudritzki}, {Magnier},
  {Martin}, {Metcalfe}, {Tonry}, {Wainscoat}, \& {Waters}}]{Green2014}
{Green}, G.~M., {et~al.} 2014, \apj, 783, 114

\bibitem[{{Green} {et~al.}(2015){Green}, {Schlafly}, {Finkbeiner}, {Rix},
  {Martin}, {Burgett}, {Draper}, {Flewelling}, {Hodapp}, {Kaiser}, {Kudritzki},
  {Magnier}, {Metcalfe}, {Price}, {Tonry}, \& {Wainscoat}}]{Green2015}
{Green}, G.~M., {et~al.} 2015, \apj, 810, 25

\bibitem[{{Guseinov} {et~al.}(2004){Guseinov}, {Ankay}, \&
  {Tagieva}}]{Guseinov2004}
{Guseinov}, O.~H., {Ankay}, A., \& {Tagieva}, S.~O. 2004, Serbian Astronomical
  Journal, 168

\bibitem[{{Hanson} {et~al.}(2016){Hanson}, {Bailer-Jones}, {Burgett},
  {Chambers}, {Hodapp}, {Kaiser}, {Tonry}, {Wainscoat}, \&
  {Waters}}]{Hanson2016}
{Hanson}, R.~J., {et~al.} 2016, \mnras, 463, 3604

\bibitem[{{Hewitt} {et~al.}(2009){Hewitt}, {Rho}, {Andersen}, \&
  {Reach}}]{Hewitt2009}
{Hewitt}, J.~W., {Rho}, J., {Andersen}, M., \& {Reach}, W.~T. 2009, \apj, 694,
  1266

\bibitem[{{Huang} \& {Thaddeus}(1986)}]{Huang1986}
{Huang}, Y.-L. \& {Thaddeus}, P. 1986, \apj, 309, 804

\bibitem[{{Ivezi{\'c}} {et~al.}(2008){Ivezi{\'c}}, {Sesar}, {Juri{\'c}},
  {Bond}, {Dalcanton}, {Rockosi}, {Yanny}, {Newberg}, {Beers}, {Allende
  Prieto}, {Wilhelm}, {Lee}, {Sivarani}, {Norris}, {Bailer-Jones}, {Re
  Fiorentin}, {Schlegel}, {Uomoto}, {Lupton}, {Knapp}, {Gunn}, {Covey},
  {Smith}, {Miknaitis}, {Doi}, {Tanaka}, {Fukugita}, {Kent}, {Finkbeiner},
  {Munn}, {Pier}, {Quinn}, {Hawley}, {Anderson}, {Kiuchi}, {Chen}, {Bushong},
  {Sohi}, {Haggard}, {Kimball}, {Barentine}, {Brewington}, {Harvanek},
  {Kleinman}, {Krzesinski}, {Long}, {Nitta}, {Snedden}, {Lee}, {Harris},
  {Brinkmann}, {Schneider}, \& {York}}]{Ivezic2008}
{Ivezi{\'c}}, {\v Z}., {et~al.} 2008, \apj, 684, 287

\bibitem[{{Jeong} {et~al.}(2012){Jeong}, {Byun}, {Koo}, {Lee}, {Lee}, \&
  {Kang}}]{Jeong2012}
{Jeong}, I.-G., {Byun}, D.-Y., {Koo}, B.-C., {Lee}, J.-J., {Lee}, J.-W., \&
  {Kang}, H. 2012, \apss, 342, 389

\bibitem[{Jiang {et~al.}(2010)Jiang, Chen, Wang, Su, Zhou, Safi-Harb, \&
  DeLaney}]{Jiang2010}
Jiang, B., Chen, Y., Wang, J., Su, Y., Zhou, X., Safi-Harb, S., \& DeLaney, T.
  2010, The Astrophysical Journal, 712, 1147

\bibitem[{{Kaiser} {et~al.}(2002){Kaiser}, {Aussel}, {Burke}, {Boesgaard},
  {Chambers}, {Chun}, {Heasley}, {Hodapp}, {Hunt}, {Jedicke}, {Jewitt},
  {Kudritzki}, {Luppino}, {Maberry}, {Magnier}, {Monet}, {Onaka}, {Pickles},
  {Rhoads}, {Simon}, {Szalay}, {Szapudi}, {Tholen}, {Tonry}, {Waterson}, \&
  {Wick}}]{Kaiser2002}
{Kaiser}, N., {et~al.} 2002, in Society of Photo-Optical Instrumentation
  Engineers (SPIE) Conference Series, Vol. 4836, Survey and Other Telescope
  Technologies and Discoveries, ed. J.~A. {Tyson} \& S.~{Wolff}, 154--164

\bibitem[{{Katsuta} {et~al.}(2012){Katsuta}, {Uchiyama}, {Tanaka}, {Tajima},
  {Bechtol}, {Funk}, {Lande}, {Ballet}, {Hanabata}, {Lemoine-Goumard}, \&
  {Takahashi}}]{Katsuta2012}
{Katsuta}, J., {et~al.} 2012, \apj, 752, 135

\bibitem[{{Kirkpatrick} {et~al.}(2014){Kirkpatrick}, {Schneider},
  {Fajardo-Acosta}, {Gelino}, {Mace}, {Wright}, {Logsdon}, {McLean}, {Cushing},
  {Skrutskie}, {Eisenhardt}, {Stern}, {Balokovi{\'c}}, {Burgasser}, {Faherty},
  {Lansbury}, {Rich}, {Skrzypek}, {Fowler}, {Cutri}, {Masci}, {Conrow},
  {Grillmair}, {McCallon}, {Beichman}, \& {Marsh}}]{Kirkpatrick2014}
{Kirkpatrick}, J.~D., {et~al.} 2014, \apj, 783, 122

\bibitem[{{Kirshner} \& {Arnold}(1979)}]{Kirshner1979}
{Kirshner}, R.~P. \& {Arnold}, C.~N. 1979, \apj, 229, 147

\bibitem[{{Kramer} {et~al.}(2003){Kramer}, {Lyne}, {Hobbs}, {L{\"o}hmer},
  {Carr}, {Jordan}, \& {Wolszczan}}]{Kramer2003}
{Kramer}, M., {Lyne}, A.~G., {Hobbs}, G., {L{\"o}hmer}, O., {Carr}, P.,
  {Jordan}, C., \& {Wolszczan}, A. 2003, \apjl, 593, L31

\bibitem[{{Kundu} {et~al.}(1980){Kundu}, {Angerhofer}, {Fuerst}, \&
  {Hirth}}]{Kundu1980}
{Kundu}, M.~R., {Angerhofer}, P.~E., {Fuerst}, E., \& {Hirth}, W. 1980, \aap,
  92, 225

\bibitem[{{Lallement} {et~al.}(2014){Lallement}, {Vergely}, {Valette},
  {Puspitarini}, {Eyer}, \& {Casagrande}}]{Lallement2014}
{Lallement}, R., {Vergely}, J.-L., {Valette}, B., {Puspitarini}, L., {Eyer},
  L., \& {Casagrande}, L. 2014, \aap, 561, A91

\bibitem[{{Liu} {et~al.}(2014){Liu}, {Yuan}, {Huo}, {Deng}, {Hou}, {Zhao},
  {Zhao}, {Shi}, {Luo}, {Xiang}, {Zhang}, {Huang}, \& {Zhang}}]{Liu2014}
{Liu}, X.-W., {et~al.} 2014, in IAU Symposium, Vol. 298, IAU Symposium, ed.
  S.~{Feltzing}, G.~{Zhao}, N.~A. {Walton}, \& P.~{Whitelock}, 310--321

\bibitem[{{Liu} {et~al.}(2015){Liu}, {Zhao}, \& {Hou}}]{Liu2015}
{Liu}, X.-W., {Zhao}, G., \& {Hou}, J.-L. 2015, Research in Astronomy and
  Astrophysics, 15, 1089

\bibitem[{{Lombardi} {et~al.}(2011){Lombardi}, {Alves}, \&
  {Lada}}]{Lombardi2011}
{Lombardi}, M., {Alves}, J., \& {Lada}, C.~J. 2011, \aap, 535, A16

\bibitem[{{Ng} {et~al.}(2007){Ng}, {Romani}, {Brisken}, {Chatterjee}, \&
  {Kramer}}]{Ng2007}
{Ng}, C.-Y., {Romani}, R.~W., {Brisken}, W.~F., {Chatterjee}, S., \& {Kramer},
  M. 2007, \apj, 654, 487

\bibitem[{{Phillips} {et~al.}(1981){Phillips}, {Gondhalekar}, \&
  {Blades}}]{Phillips1981}
{Phillips}, A.~P., {Gondhalekar}, P.~M., \& {Blades}, J.~C. 1981, \mnras, 195,
  485

\bibitem[{{Planck Collaboration} {et~al.}(2014){Planck Collaboration},
  {Abergel}, {Ade}, {Aghanim}, {Alves}, {Aniano}, {Armitage-Caplan}, {Arnaud},
  {Ashdown}, {Atrio-Barandela}, \& et~al.}]{Planck2014}
{Planck Collaboration}, {et~al.} 2014, \aap, 571, A11

\bibitem[{{Planck Collaboration} {et~al.}(2011){Planck Collaboration}, {Ade},
  {Aghanim}, {Arnaud}, {Ashdown}, {Aumont}, {Baccigalupi}, {Balbi}, {Banday},
  {Barreiro}, \& et~al.}]{Planck2011}
{Planck Collaboration}, {et~al.} 2011, \aap, 536, A19

\bibitem[{{Sallmen} \& {Welsh}(2004)}]{Sallmen2004}
{Sallmen}, S. \& {Welsh}, B.~Y. 2004, \aap, 426, 555

\bibitem[{{Sauvageot} {et~al.}(1990){Sauvageot}, {Ballet}, \&
  {Rothenflug}}]{Sauvageot1990}
{Sauvageot}, J.~L., {Ballet}, J., \& {Rothenflug}, R. 1990, \aap, 227, 183

\bibitem[{{Schlafly} \& {Finkbeiner}(2011)}]{Schlafly2011}
{Schlafly}, E.~F. \& {Finkbeiner}, D.~P. 2011, \apj, 737, 103

\bibitem[{{Schlafly} {et~al.}(2014){Schlafly}, {Green}, {Finkbeiner}, {Rix},
  {Bell}, {Burgett}, {Chambers}, {Draper}, {Hodapp}, {Kaiser}, {Magnier},
  {Martin}, {Metcalfe}, {Price}, \& {Tonry}}]{Schlafly2014}
{Schlafly}, E.~F., {et~al.} 2014, \apj, 786, 29

\bibitem[{Schlafly {et~al.}(2015)Schlafly, Green, Finkbeiner, Rix, Burgett,
  Chambers, Draper, Kaiser, Martin, Metcalfe, Morgan, Price, Tonry, Wainscoat,
  \& Waters}]{Schlafly2015}
Schlafly, E.~F., {et~al.} 2015, \apj, 799, 116

\bibitem[{{Schlegel} {et~al.}(1998){Schlegel}, {Finkbeiner}, \&
  {Davis}}]{Schlegel1998}
{Schlegel}, D.~J., {Finkbeiner}, D.~P., \& {Davis}, M. 1998, \apj, 500, 525

\bibitem[{{Schultheis} {et~al.}(2014){Schultheis}, {Chen}, {Jiang}, {Gonzalez},
  {Enokiya}, {Fukui}, {Torii}, {Rejkuba}, \& {Minniti}}]{Schultheis2014}
{Schultheis}, M., {et~al.} 2014, \aap, 566, A120

\bibitem[{{Seta} {et~al.}(1998){Seta}, {Hasegawa}, {Dame}, {Sakamoto}, {Oka},
  {Handa}, {Hayashi}, {Morino}, {Sorai}, \& {Usuda}}]{Seta1998}
{Seta}, M., {et~al.} 1998, \apj, 505, 286

\bibitem[{{Skrutskie} {et~al.}(2006){Skrutskie}, {Cutri}, {Stiening},
  {Weinberg}, {Schneider}, {Carpenter}, {Beichman}, {Capps}, {Chester},
  {Elias}, {Huchra}, {Liebert}, {Lonsdale}, {Monet}, {Price}, {Seitzer},
  {Jarrett}, {Kirkpatrick}, {Gizis}, {Howard}, {Evans}, {Fowler}, {Fullmer},
  {Hurt}, {Light}, {Kopan}, {Marsh}, {McCallon}, {Tam}, {Van Dyk}, \&
  {Wheelock}}]{Skrutskie2006}
{Skrutskie}, M.~F., {et~al.} 2006, \aj, 131, 1163

\bibitem[{{Sofue} {et~al.}(1980){Sofue}, {Furst}, \& {Hirth}}]{Sofue1980}
{Sofue}, Y., {Furst}, E., \& {Hirth}, W. 1980, \pasj, 32, 1

\bibitem[{{Sun} {et~al.}(1996){Sun}, {Anderson}, {Aschenbach}, {Becker},
  {Egger}, {Kanbach}, {Merck}, {Tr{\"u}mper}, \& {Wolszczan}}]{Sun1996}
{Sun}, X., {et~al.} 1996, in Roentgenstrahlung from the Universe, ed. H.~U.
  {Zimmermann}, J.~{Tr{\"u}mper}, \& H.~{Yorke}, 195--196

\bibitem[{{van den Bergh} {et~al.}(1973){van den Bergh}, {Marscher}, \&
  {Terzian}}]{vandenBergh1973}
{van den Bergh}, S., {Marscher}, A.~P., \& {Terzian}, Y. 1973, \apjs, 26, 19

\bibitem[{{Voges} {et~al.}(1999){Voges}, {Aschenbach}, {Boller},
  {Br{\"a}uninger}, {Briel}, {Burkert}, {Dennerl}, {Englhauser}, {Gruber},
  {Haberl}, {Hartner}, {Hasinger}, {K{\"u}rster}, {Pfeffermann}, {Pietsch},
  {Predehl}, {Rosso}, {Schmitt}, {Tr{\"u}mper}, \& {Zimmermann}}]{Voges1999}
{Voges}, W., {et~al.} 1999, \aap, 349, 389

\bibitem[{{Wright} {et~al.}(2010){Wright}, {Eisenhardt}, {Mainzer}, {Ressler},
  {Cutri}, {Jarrett}, {Kirkpatrick}, {Padgett}, {McMillan}, {Skrutskie},
  {Stanford}, {Cohen}, {Walker}, {Mather}, {Leisawitz}, {Gautier}, {McLean},
  {Benford}, {Lonsdale}, {Blain}, {Mendez}, {Irace}, {Duval}, {Liu}, {Royer},
  {Heinrichsen}, {Howard}, {Shannon}, {Kendall}, {Walsh}, {Larsen}, {Cardon},
  {Schick}, {Schwalm}, {Abid}, {Fabinsky}, {Naes}, \& {Tsai}}]{Wright2010}
{Wright}, E.~L., {et~al.} 2010, \aj, 140, 1868

\bibitem[{{Xiang} {et~al.}(2017{\natexlab{a}}){Xiang}, {Liu}, {Shi}, {Yuan},
  {Huang}, {Luo}, {Zhang}, {Zhao}, {Zhang}, {Ren}, {Chen}, {Wang}, {Li}, {Huo},
  {Zhang}, {Wang}, {Zhang}, {Hou}, \& {Wang}}]{Xiang2016b}
{Xiang}, M.-S., {et~al.} 2017{\natexlab{a}}, \mnras, 464, 3657

\bibitem[{{Xiang} {et~al.}(2017{\natexlab{b}}){Xiang}, {Liu}, {Shi}, {Yuan},
  {Huang}, {Luo}, {Zhang}, {Zhao}, {Zhang}, {Ren}, {Chen}, {Wang}, {Li}, {Huo},
  {Zhang}, {Wang}, {Zhang}, {Hou}, \& {Wang}}]{Xiang2016}
{Xiang}, M.-S., {et~al.} 2017{\natexlab{b}}, \mnras, submited

\bibitem[{{Xiao} {et~al.}(2008){Xiao}, {F{\"u}rst}, {Reich}, \&
  {Han}}]{Xiao2008}
{Xiao}, L., {F{\"u}rst}, E., {Reich}, W., \& {Han}, J.~L. 2008, \aap, 482, 783

\bibitem[{{Yuan} {et~al.}(2015){Yuan}, {Liu}, {Huo}, {Xiang}, {Huang}, {Chen},
  {Zhang}, {Sun}, {Wang}, {Zhang}, {Zhao}, {Luo}, {Shi}, {Li}, {Yuan}, {Dong},
  {Li}, {Hou}, \& {Zhang}}]{Yuan2015}
{Yuan}, H.-B., {et~al.} 2015, \mnras, 448, 855

\bibitem[{{Yuan} {et~al.}(2013){Yuan}, {Liu}, \& {Xiang}}]{Yuan2013}
{Yuan}, H.~B., {Liu}, X.~W., \& {Xiang}, M.~S. 2013, \mnras, 430, 2188

\bibitem[{{Zhang} {et~al.}(2013){Zhang}, {Liu}, {Yuan}, {Zhao}, {Yao}, {Zhang},
  \& {Xiang}}]{Zhang2013}
{Zhang}, H.-H., {Liu}, X.-W., {Yuan}, H.-B., {Zhao}, H.-B., {Yao}, J.-S.,
  {Zhang}, H.-W., \& {Xiang}, M.-S. 2013, Research in Astronomy and
  Astrophysics, 13, 490

\bibitem[{{Zhang} {et~al.}(2014){Zhang}, {Liu}, {Yuan}, {Zhao}, {Yao}, {Zhang},
  \& {Huang}}]{Zhang2014}
{Zhang}, H.-H., {Liu}, X.-W., {Yuan}, H.-B., {Zhao}, H.-B., {Yao}, J.-S.,
  {Zhang}, H.-W.~{Xiang}, M.-S., \& {Huang}, Y. 2014, Research in Astronomy and
  Astrophysics, 14, 456

\end{thebibliography}

\label{lastpage}
\end{document}